\documentclass[journal]{IEEEtran}
\usepackage{amssymb}
\usepackage{graphicx}
\usepackage{amsmath}
\usepackage[nooneline,flushleft]{caption2}
\usepackage{subfigure}
\usepackage{multirow}
\usepackage{multicol}
\usepackage{stfloats}
\usepackage{color}
\usepackage{algorithm}
\usepackage{algpseudocode}
\usepackage{epstopdf}
\usepackage{tikz}

\usepackage{caption2}

\usepackage{amsthm}

\DeclareMathOperator*{\argmin}{argmin}

\theoremstyle{proof}

\renewcommand\arraystretch{1.5}
\usepackage{makecell}

\begin{document}

\title{Rate-Distortion Modeling for Bit Rate Constrained Point Cloud Compression}
\author{Pan Gao, Shengzhou Luo, and Manoranjan Paul
\thanks{Copyright © 20xx IEEE. Personal use of this material is permitted. However, permission to use this material for any other purposes must be obtained from the IEEE by sending an email to pubs-permissions@ieee.org.}
\thanks{This work was supported in part by the National Key Research and Development Program of China under Grant 2021ZD0113203 and the Natural Science Foundation of China under Grant 62272227.}
\thanks{P. Gao is  with College of Computer Science and Technology, Nanjing University of Aeronautics and Astronautics, Nanjing 211106, P. R. China, and  also with MIIT Key Laboratory of Pattern Analysis and Machine Intelligence, Nanjing University of Aeronautics and Astronautics, Nanjing 211106, P. R. China (Email: gaopan.1005@gmail.com).  
	
	S. Luo is with School of Software, South China Normal University, Foshan 528225, P. R. China (Email: luos@m.scnu.edu.cn). 
	
	M. Paul is with School of Computing and Mathematics, Charles Sturt University, Bathurst,  NSW 2795, Australia (Email: mpaul@csu.edu.au).}
}

\maketitle

\begin{abstract}
As being one of the main representation formats of 3D real world and well-suited for virtual reality and augmented reality applications, point clouds have gained a lot of popularity. In order to reduce the huge amount of data, a considerable amount of research on point cloud compression has been done. However, given a target bit rate, how to properly choose the color and geometry quantization parameters for compressing point clouds is still an open issue. In this paper, we propose a rate-distortion model based quantization parameter selection scheme for bit rate constrained point cloud compression. Firstly, to overcome the measurement uncertainty in evaluating the distortion of the point clouds, we propose a unified model to combine the geometry distortion and color distortion. In this model, we take into account the correlation between geometry and color variables of point clouds and derive a dimensionless quantity to represent the overall quality degradation. Then, we derive the relationships of overall distortion and bit rate with the quantization parameters. Finally, we formulate the bit rate constrained point cloud compression as a constrained minimization problem using the derived polynomial models and deduce the solution via an iterative numerical method.  Experimental results show that the proposed algorithm can achieve optimal decoded point cloud quality at various target bit rates, and substantially outperform the \emph{video-rate-distortion model} based point cloud compression scheme.  

\end{abstract}


\section{Introduction}
Due to the advancement in 3D scene capturing and rendering, point clouds have been a popular representation in emerging applications, such as metaverse, 3D telepresence, gaming, and virtual/augmented reality \cite{3D_point_clouds_AR, 6DoF_video, Point_Cloud_clutering}. Recently, compression of point clouds has gained a significant attention from both academia and industry \cite{acmm2019, ACM_web3d} to enable the transmission of the captured dynamic 3D scenes or objects to a remote location. However, almost all the previous efforts on point cloud compression are focused on addressing the problem of how to reduce the bit rate given a reconstruction quality level (i.e., quantization parameter, QP), or improve the reconstruction quality at the same bit rate. Given a bit rate constraint in the bandwidth limited channel, finding the optimal QPs that can achieve maximal reconstruction quality is still an open issue in point cloud compression. 

In traditional video compression and optimization, bit rate constrained encoder parameter optimization is formulated and solved as follows. Assume that there is a set of QPs to be determined, which is represented as ${\rm\bf{q}}=[q_1,\cdots,q_N]$, where $N$ is the number of QPs. The problem of finding the optimal QP vector  $\rm\bf{q^*}$ subject to bit rate constraint $\widehat R$ can be usually expressed as
\begin{equation}
\begin{aligned}
& {\rm\bf{q^*}}=\operatorname*{arg\,min}_{\bf{q}} D(\bf{q})\\
&  \text{s. t.} \;
  R({\bf{q}}) \leq \widehat R \\
\end{aligned}
\label{constrained_compression_formula}
\end{equation}
where $D(\bf{q})$ and $R(\bf{q})$ denote the distortion and bit rate for a particular choice of $\rm\bf{q}$, respectively. 
{Inversely, the rate-distortion optimization problem can also be formulated as the minimization of the rate given a constraint on allowable distortion.} 
In H.264/AVC and HEVC encoders, $D(\bf{q})$ and $R(\bf{q})$ are generally modeled as linear and quadratic functions of $\rm\bf{q}$, respectively \cite{rate_control_linbin}, \cite{AVC_rate_control}. 
To relax the constraint, a Lagrangian cost function is generated by combining the distortion function $D(\bf{q})$ with the bit rate constraint function $R({\bf{q}})$ multiplied by a  Lagrange multiplier $\lambda$, and then the optimal $\rm\bf{q^*}$ can be derived by minimizing the Lagrangian cost function, i.e., 
\begin{equation}
\label{Lagrangian_cost_function}
{\rm\bf{q^*}}=\operatorname*{arg\,min}_{\bf{q}} J({\bf{q}},\lambda)=\operatorname*{arg\,min}_{{\bf{q}}} D({\bf{q}})+\lambda (R({\bf{q}})-\widehat R)
\end{equation}

In \eqref{Lagrangian_cost_function}, $J(\bf{q},\lambda)$ is used to represent the unconstrained Lagrangian function. Since the Lagrangian function formulated for normal video coding is convex, to find the optimal solution of \eqref{Lagrangian_cost_function}, the following optimality conditions are employed,
\begin{equation}\label{gradient_with_q}
\begin{aligned}
&\nabla_{\bf{q}} J({\bf{q}},\lambda)= \nabla_{\bf{q}} D({\bf{q}}) + \lambda \nabla_{\bf{q}} R({\bf{q}})=0\\
&\nabla_{\lambda} J({\bf{q}},\lambda)=R({\bf{q}})-\widehat R=0
\end{aligned}
\end{equation}

In \eqref{gradient_with_q}, the gradients of the $J(\bf{q},\lambda)$ with the $\rm\bf{q}$ and $\lambda$ are derived, and both  are equal to zero when the cost function achieves the minimum value. With these gradients,  $N+1$ equations are obtained for the $N+1$ unknown variables, and thus the optimal solution can be determined and solved, in which a closed-form solution can usually exist due to uncomplicated rate and distortion models in traditional video compression.  

When considered for point cloud compression, bit rate constrained encoder parameter optimization faces major new challenges. Firstly, different from the normal video, a point in the point clouds have geometry position and color attribute information. Therefore,  two associated QPs, ${\rm\bf{q}}=[q_g,q_c]$,  are needed for compressing point clouds, and then there will be two different kinds of compression distortions generated, i.e., $D(q_g)$ and $D(q_c)$. However, there is still no a unified model $f$ in the literature that can properly combine these two distortions, i.e., making $D({\bf{q}})=f(D(q_g)+D(q_c))$. Nevertheless, a unified model is always necessary for point cloud optimization. This is because that, as geometry distortion $D(q_g)$ and color distortion $D(q_c)$ are measured at different scales, we need an overall model to balance these two distortions. Further, the color distortion $D(q_c)$ for each point is calculated using the color difference between the point in the compressed point cloud and the matching point in the original point cloud identified by the associated geometry Euclidean distance. Thus, the color distortion $D(q_c)$ is dependent on $D(q_g)$, and   $D(q_c)$ cannot be separated from $D(q_g)$ in the measurement of compression distortion for point clouds. 
Secondly, it has still remained unclear that what the relationships of  $D(\bf{q})$ with $q_g$ and $q_c$ are. In point cloud compression, the number of points of the reconstructed point clouds is usually different from that of the original point clouds. That means, many-to-many mapping exists for the points. This type of point correspondence may induce that the classic linear and quadratic models are no longer applicable, and high degree polynomials may arise for characterizing the effect of quantization on bit rate and distortion.
Finally, due to possible more complicated rate distortion models involved, there may not exist closed-form solution for the  Lagrangian-formulated unconstrained function. How to relax the bit rate constraint and find the corresponding optimal solution is thus another difficulty in point cloud optimization. 

In this paper, we propose a rate-distortion optimization scheme for finding the optimal QPs for bit rate constrained point cloud compression. Firstly, we propose a unified model to combine the geometry distortion and color distortion, in which distortion measurement uncertainty induced by different scales used on the geometry and color variables is overcome by using the covariance matrix of the point clouds.  Then, with the unified distortion model, we derive  polynomial models to model the relationship of the overall distortion and bit rate with geometry and color QPs. Finally, a model-based bit rate allocation scheme is formulated to allocate the bit rate for geometry and color coding under a target bit rate. In this scheme, we develop an augmented Lagrangian method to solve the constrained bit rate allocation problem. 

It should be noted here that, in this paper, we aim at finding the optimal geometry QP and color QP for a given bit rate. Currently, there are two types of point cloud compression frameworks, i.e., geometry based point cloud compression (G-PCC) and video based point cloud compression (V-PCC) \cite{Comprehensive_PCC}. In G-PCC, geometry coding is performed by octree coding, and lossy geometry coding is primarily enabled by grouping the leaf nodes (or voxels) into blocks at a higher level or directly using position scaling \cite{volumetric_approach}. Thus, bit rate allocation between geometry and color in G-PCC can be recasted into a problem of search of the octree level of geometry and the QP of color.  Since the octree levels (or scaling factors) are rather limited, this problem can be addressed by using the point cloud sub-sampling strategy in a tractable way \cite{ACM_PV}.  In contrast, in V-PCC, both geometry QP and color QP are utilized for compressing point clouds, and geometry distortion and color distortion are combined in a non-linear fashion to affect the overall quality of point clouds. Therefore, compared to G-PCC, optimization of compression parameters of V-PCC is more challenging. To this end,  we focus on rate distortion modeling for video-based point cloud compression in this work. 


The rest of the paper is organized as follows. We review the recent works on point cloud compression in Section II. The rate and distortion models specific for point cloud compression are derived in Section III, in which a unified distortion model is developed to represent the overall quality degradation of point clouds. We perform rate distortion optimization for bit rate constrained point cloud coding in Section \ref{RDO_PCC}. Experiments are conducted in Section \ref{Experiments}, and concluding remarks are given in Section \ref{conclusion}.

\section{Related Work}
Compared to traditional polygonal mesh representations of 3D object requiring the connectivity to specify the surface topology,   point cloud representation only contains the geometry locations of the object and the attributes (e.g., color), which can provide more flexible rendering  capability \cite{Pereira, JESTCAS}. Usually, a point cloud frame contains millions of points, which poses a large amount of data. Thus, how to efficiently compress the point clouds becomes a central component in virtual/augmented reality applications. Generally speaking, point cloud compression can be classified as two categories, i.e., static point cloud compression and dynamic point cloud compression, similar to intra-frame coding and inter-frame coding in 2-D video coding \cite{JESTCAS}. In the following, we review recent research progress in these two categories. 

\subsection{Static Point Cloud Compression}
In static point cloud compression, octree coding is the most popular way to compress the geometry information. In octree coding, the bounding cube of the point clouds is subdivided into eight child cells, and the occupied child cells are subdivided further. For each octree cell subdivision, a 8-bit occupancy code is encoded to signify which child cells are occupied. To achieve high compression efficiency for octee-based point cloud compression, Schnabel \emph{et al.} \cite{Occtree_coding} proposed to use local surface approximation to predict the number of non-empty child cells and the child cell configuration. With the same objective as in \cite{Occtree_coding}, Huang \emph{et al.} \cite{TVCG_PCC} proposed an occupancy code reordering scheme to lower the entropy of codes representing the non-empty child cells. This octree-based codec is then coupled with normal coder and color coder for the purpose of compressing the normal and color attributes, respectively. In \cite{Cha_Zhang}, by considering the fact that the points in the point clouds are unstructured,  Zhang \emph{et al.} proposed to use graph transform for point cloud attribute compression. In this approach, the graph transform basis is obtained by eigenvalue decomposition of the graph Laplacian constructed on small neighbourhoods of the point clouds. To exploit the compactness properties of graph transform, the authors of \cite{GTGC} extend the graph transform for coding the geometry information of point clouds, in which a two-layer solution is proposed. The base layer is coded with the octree-based approach, and the enhancement layer is coded with a graph-based transform approach. 
Different from the graph transform that requires generating the graph on the octree block in advance, a region adaptive hierarchical transform (RAHT) is proposed in \cite{Ricardo1}, which employs the colors of the nodes in a lower level of the octree to predict the colors of the nodes in the next level. In order to enable point cloud bit stream interoperable between 
devices and services, MPEG has started point cloud compression standardized solution. The static point cloud compression solution developed by MPEG is called G-PCC, in which the geometry coder employs octree coding and color coder uses RAHT to transform the color attributes. In the latest G-PCC model, in addition to octree coding and RAHT coding, the trisoup coding and Pred/Lift are introduced to geometry coding and color coding, respectively \cite{G-PCC-v2}.  Interested readers are referred to \cite{Geometry_PC, Geometry_PC2, G-PCC-v2} for more details.

\subsection{Dynamic Point Cloud Compression}
In dynamic point cloud compression, the temporal redundancy between consecutive point cloud frames needs to be exploited. To this end, Kammerl \emph{et al.} \cite{ICRA} proposed a double-buffering octree scheme, which encodes the structural changes within octree branch nodes using the exclusive disjunction operator. For 3D tele-immersive video applications, the authors of \cite{PCC_teleimmersive} proposed to use a rigid transform calculated from the iterative closest point algorithm to predict the points in the blocks in the predictive frame. In this scheme, the rigid transform is coded using a  quaternion quantization method. In compression of color attribute, the color per vertex is mapped to an image grid and then compressed with the JPEG standard. Analogous to traditional video compression techniques, in \cite{Ricardo2}, the authors introduced the use of motion estimation and motion compensation into dynamic point cloud compression. In this coding framework, the motion vectors of the coding blocks are inferred from the correspondences that are calculated in the 3D surface reconstruction processes. As motion compensation introduces distortion into geometry position, the authors proposed an in-loop filtering to minimize compression artifacts at the encoder. By considering the fact that the point cloud frames have varying numbers of points without explicit correspondence information, Thanou \emph{et al.} \cite{Thanou_ICIP, Thanou} developed a graph-based dynamic point cloud algorithm. In this approach, the motion vectors are firstly estimated on a sparse set of representative vertices based on spectral graph wavelet descriptors, and then, a dense motion field is interpolated. 
Similar to the standardization efforts put in static point cloud compression, MPEG has also developed compression solution for time-varying point clouds, which is called V-PCC \cite{Video_PCC, Video_PC2}.
In V-PCC, the point clouds are converted into two video sequences, i.e., texture video and geometry video, and then compressed with the existing video codecs, i.e., HEVC. For the projection video being better to comply with HEVC, the author of \cite{Li_li_tmc2} proposed to use the 3D motion and the  3D-to-2D correspondence to estimate the motion vector for texture video.  
Due to the excellent performance of HEVC in exploiting temporal correlation in video frames, V-PCC outperforms state-of-the-art dynamic point cloud compression algorithms \cite{JESTCAS, Video_PCC_2}.  

{In the area of rate distortion optimized video-based point cloud compression, considering the fact that the unoccupied pixels do not impact the reconstructed quality of the point cloud during encoding, an occupancy-map-based rate distortion optimization method is proposed \cite{RDO-occupancy}, in which a mask is used to ignore the distortion of the unoccupied pixels. {Further, an occupancy-map-based partition is proposed to divide the occupied pixels from different patches and the unoccupied pixels into different prediction units.
In  \cite{RDO-Lili}, the authors proposed a video-level bit allocation algorithm between geometry and attribute videos to optimize the overall reconstructed point cloud quality. In this scheme, the relative importance between the distortions for the geometry and attribute videos is set with a fixed value based on subjective observations. Besides the bit allocation scheme, they also proposed to assign zero bits to the unoccupied basic units, and estimate the rate model parameters of the basic units with occupied pixels from the corresponding basic units in the previous frame. }
For compression of unoccupied regions, Herglotz \emph{et al.} \cite{RDO-transform} proposed a rate-distortion optimal transform coefficient selection scheme. This scheme exploits a generalized frequency selective extrapolation approach for discrete cosine and sine transforms.  Yuan \emph{et al.} \cite{RDO-differential} tackled rate-distortion optimized video-based point cloud compression with differential evolution. To have a perceptual quality-related distortion measure in rate distortion optimization, Liu \emph{et al.} \cite{RDO-perceptual} proposed a reduced reference perceptual quality model and applied it to rate control in video-based point cloud compression. 
{The coefficients of the perceptual model are derived from two features, i.e., color fluctuation over geometric distances and color block mean variance.}   
To improve the rate-distortion performance of V-PCC, a learning-based rate control method is proposed in \cite{RDO-learning}, where a CNN-LSTM neural network is trained to predict the basic unit parameters.}
In this work, we will build our proposed bit rate constrained point cloud compression algorithm on  V-PCC. 
It should be noted that, recently, a similar joint bit allocation scheme is proposed for video based 3D point cloud compression \cite{model-based-scheme}. In this scheme, the overall distortion for point clouds is empirically characterized as the linear combination of the  geometry distortion and color distortion, and an interior point method is employed to optimize the geometry and color quantization steps after pre-encoding the point clouds. Different from this work, we theoretically derive a unified distortion model for point clouds based on the intrinsics characteristics of the  correlation between geometry variable and color attribute, and validate the unified model using a publicly available subjective database. Further, we propose an augmented Lagrangian multiplier method to find the solution to rate-constrained point cloud compression, in which the rate and distortion models are characterized as high degree polynomials.

\section{Rate-distortion modeling for point cloud compression}
In this section, we firstly build a unified distortion  model to represent the overall quality degradation of the compressed point clouds. Then, we derive the distortion and bit rate models relating to the quantization parameters by using a variety of dynamic point cloud sequences.
\begin{figure*}[!htbp]
	\hspace{-1mm}\subfigure[$q_g$ - AxeGuy]{
		\label{Fig.sub.1}
		\includegraphics[width=0.24\textwidth]{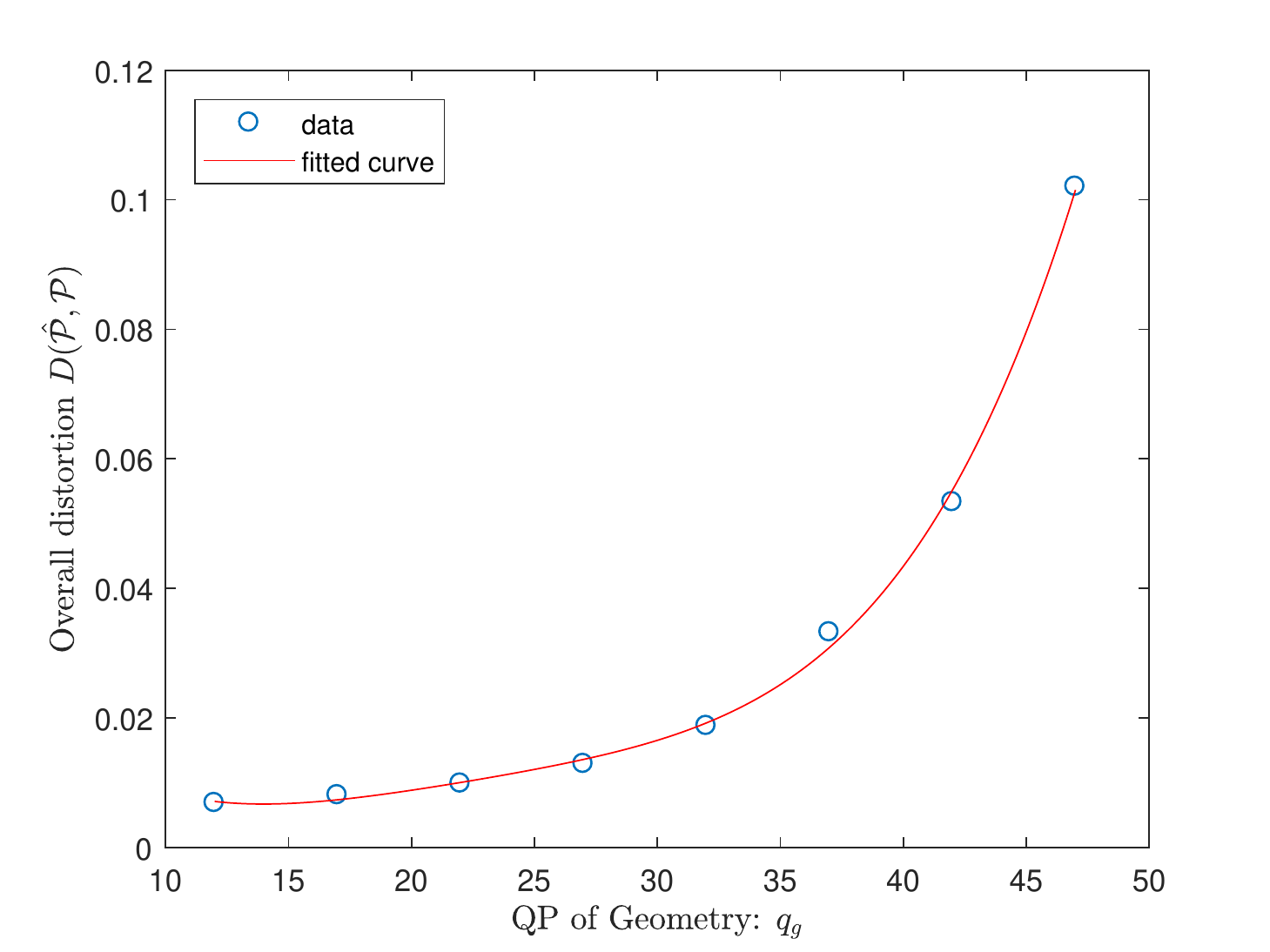}}
	\subfigure[$q_g$ - Longdress]{
		\label{Fig.sub.1}
		\includegraphics[width=0.24\textwidth]{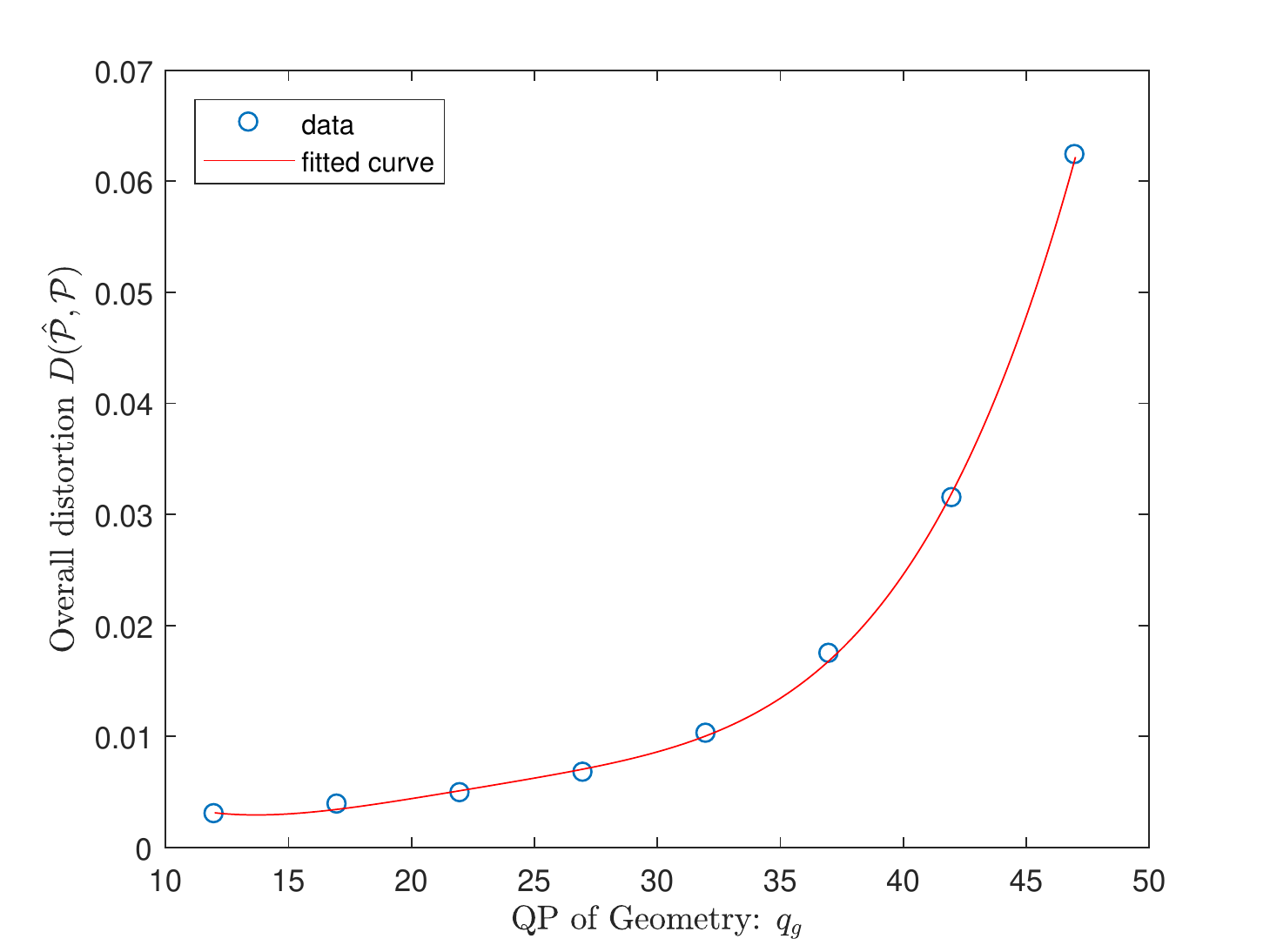}}
	\subfigure[$q_g$ - Loot]{
		\label{Fig.sub.1}
		\includegraphics[width=0.24\textwidth]{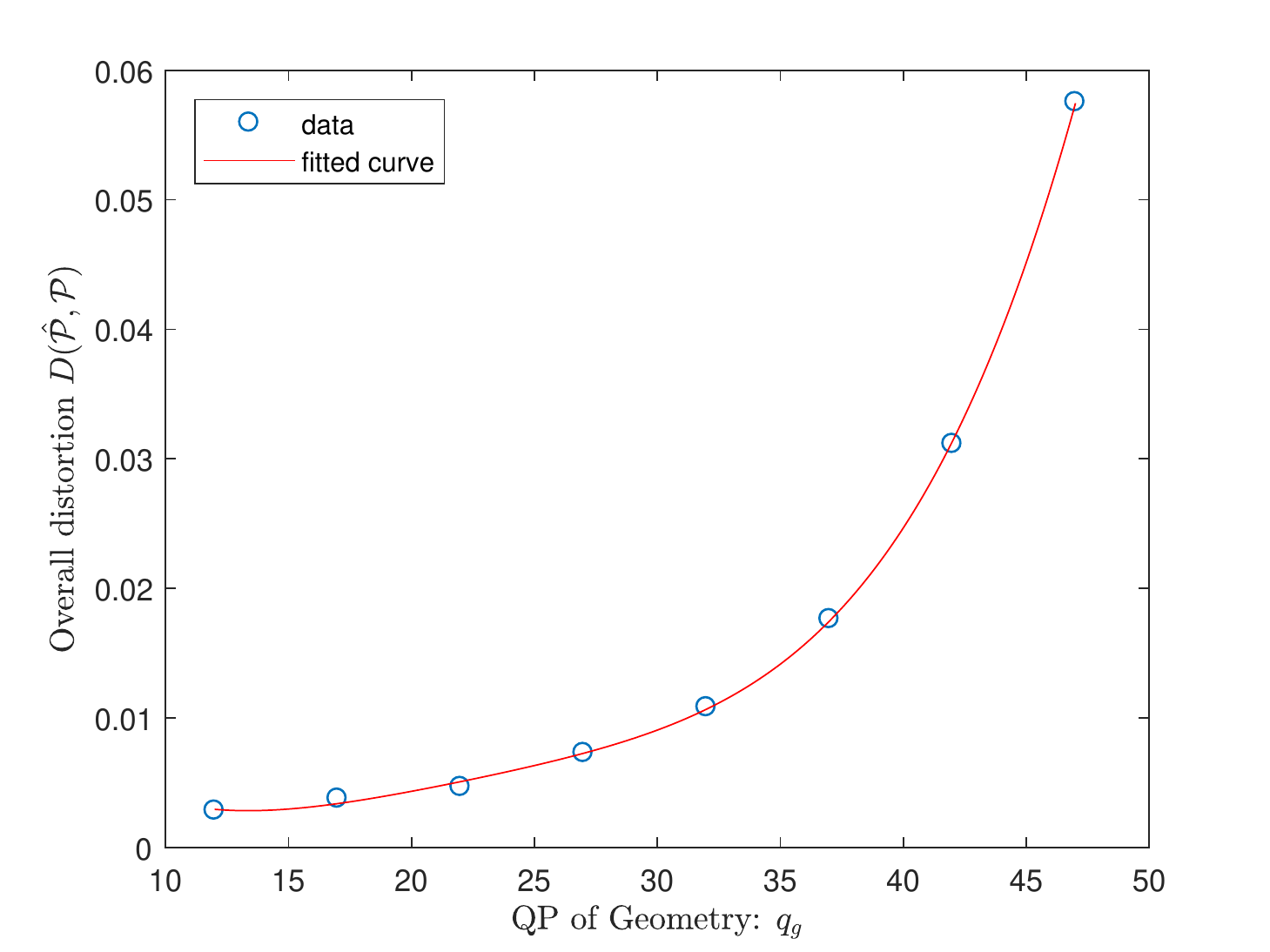}}
	\subfigure[$q_g$ - Soldier]{
		\label{Fig.sub.1}
		\includegraphics[width=0.24\textwidth]{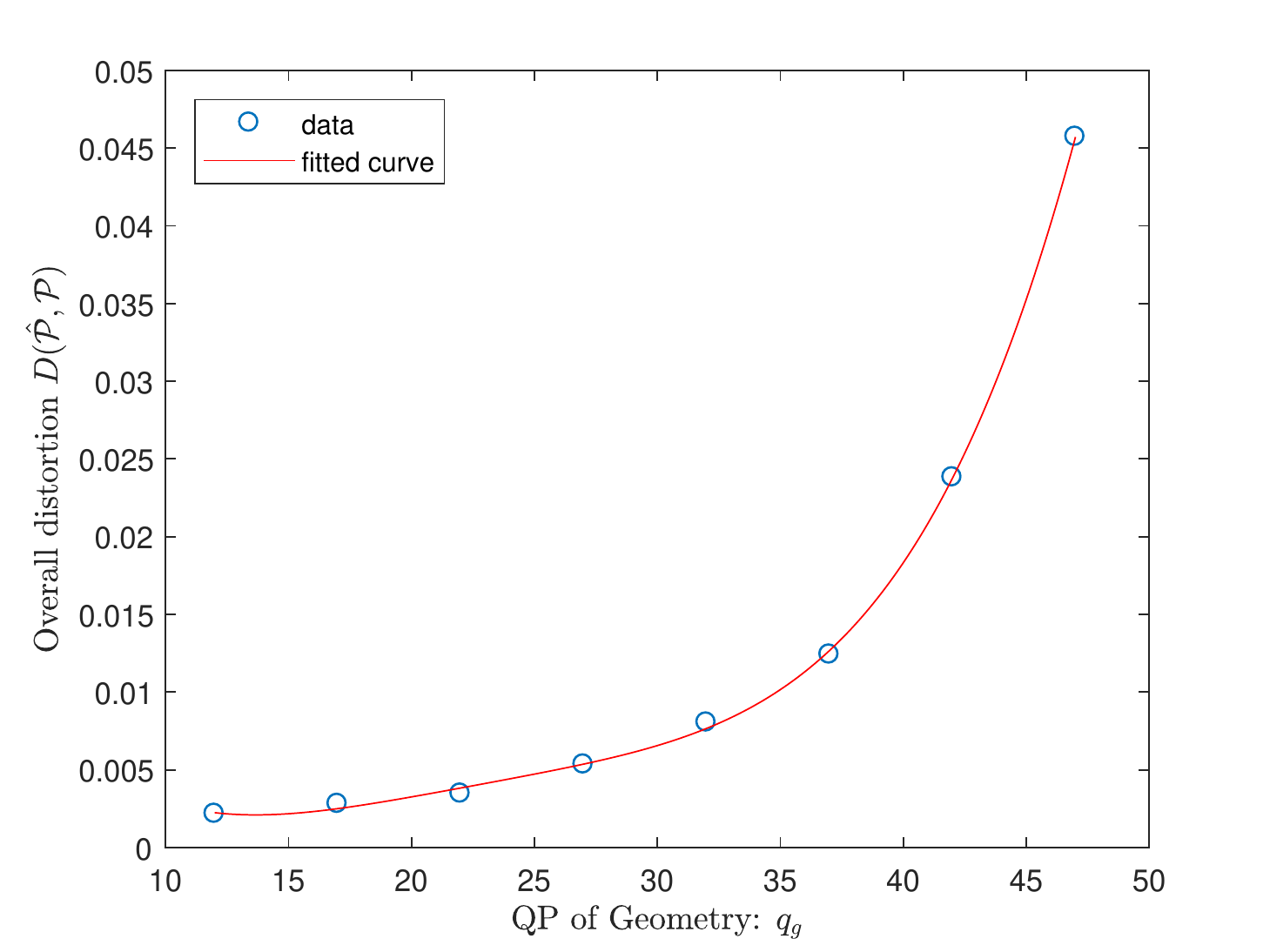}}
	
	\hspace{-1mm}\subfigure[$q_c$ - AxeGuy]{
		\label{Fig.sub.1}
		\includegraphics[width=0.24\textwidth]{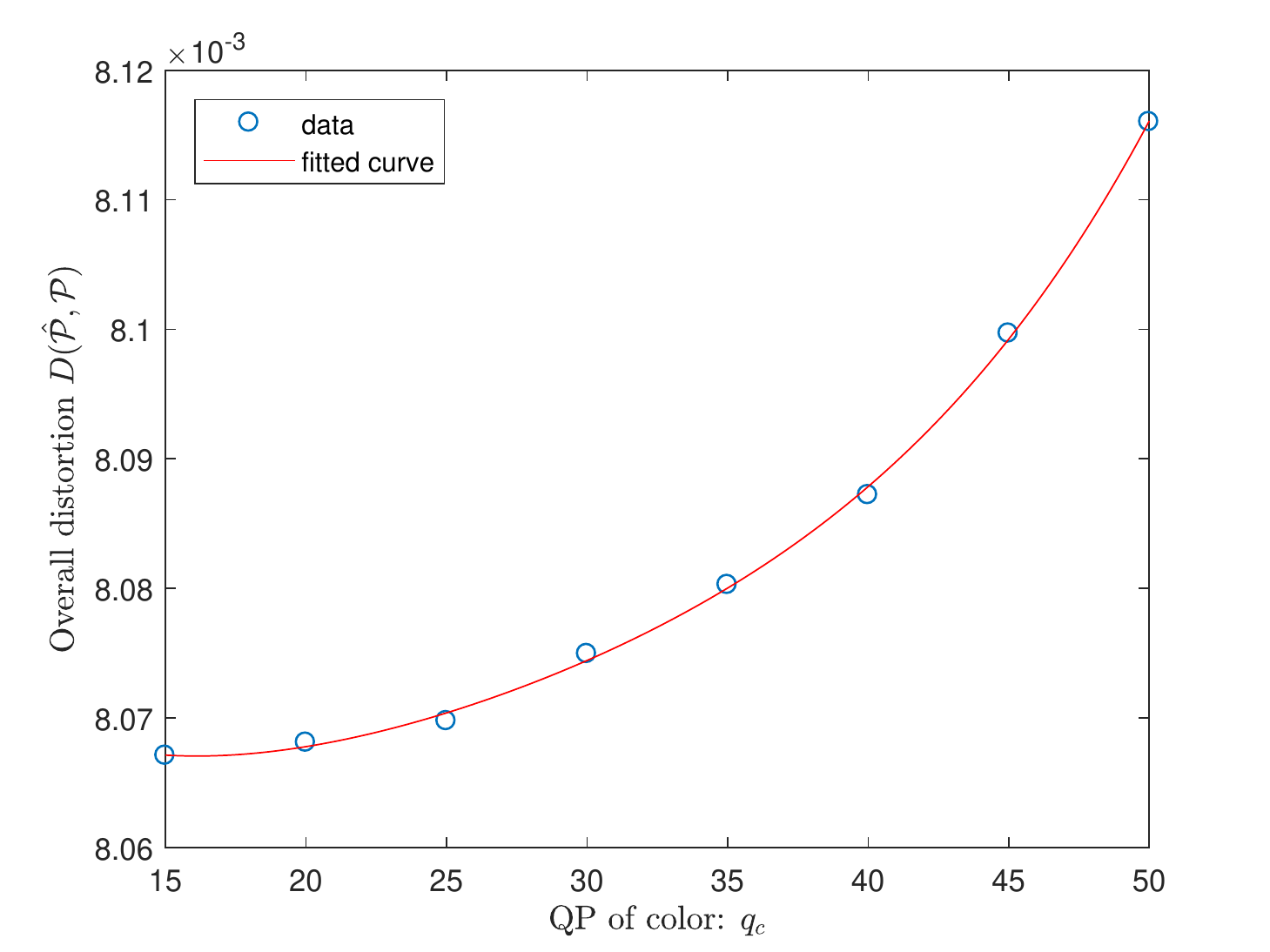}}
	\subfigure[$q_c$ - Longdress]{
		\label{Fig.sub.1}
		\includegraphics[width=0.24\textwidth]{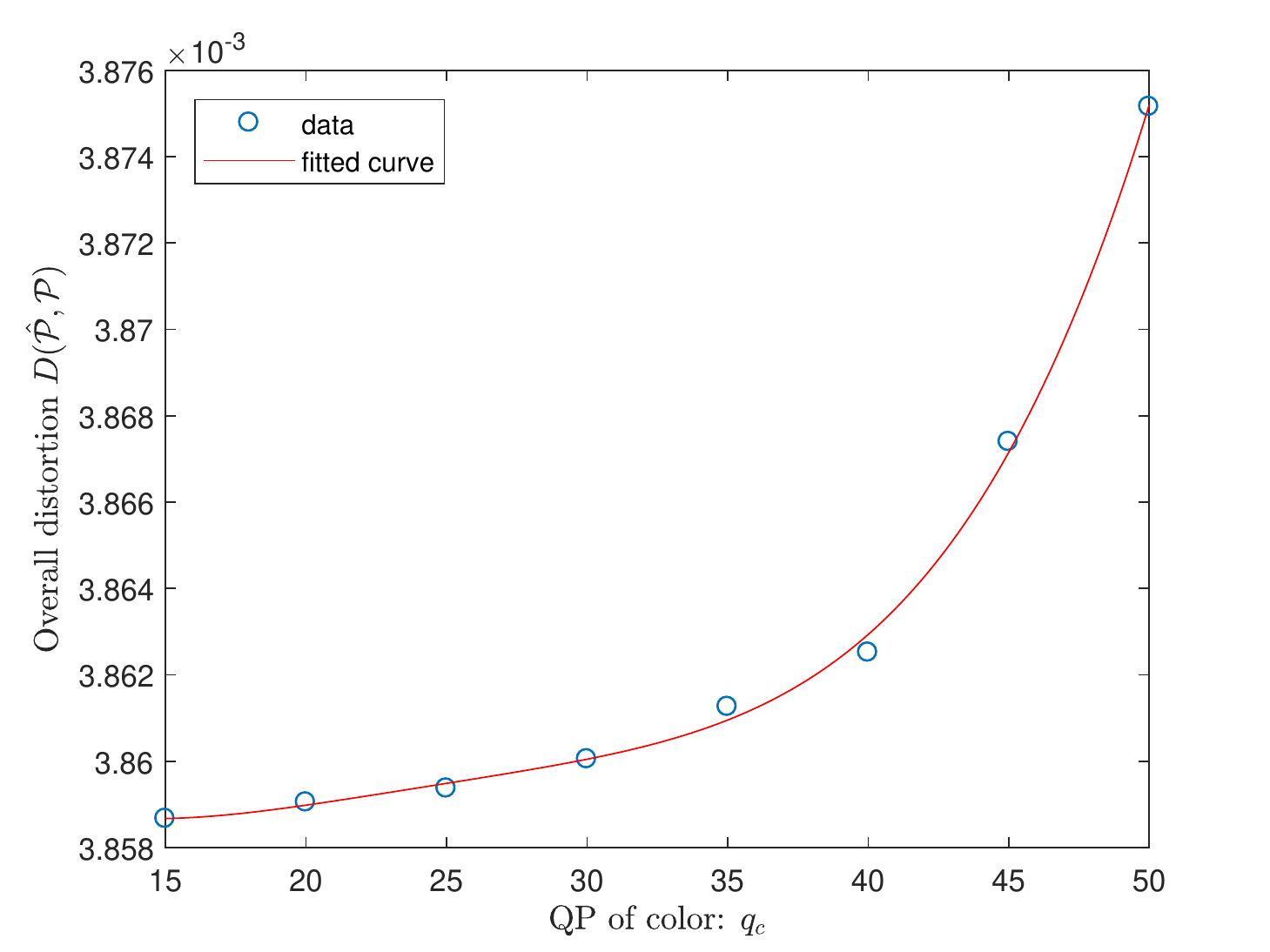}}
	\subfigure[$q_c$ - Loot]{
		\label{Fig.sub.1}
		\includegraphics[width=0.24\textwidth]{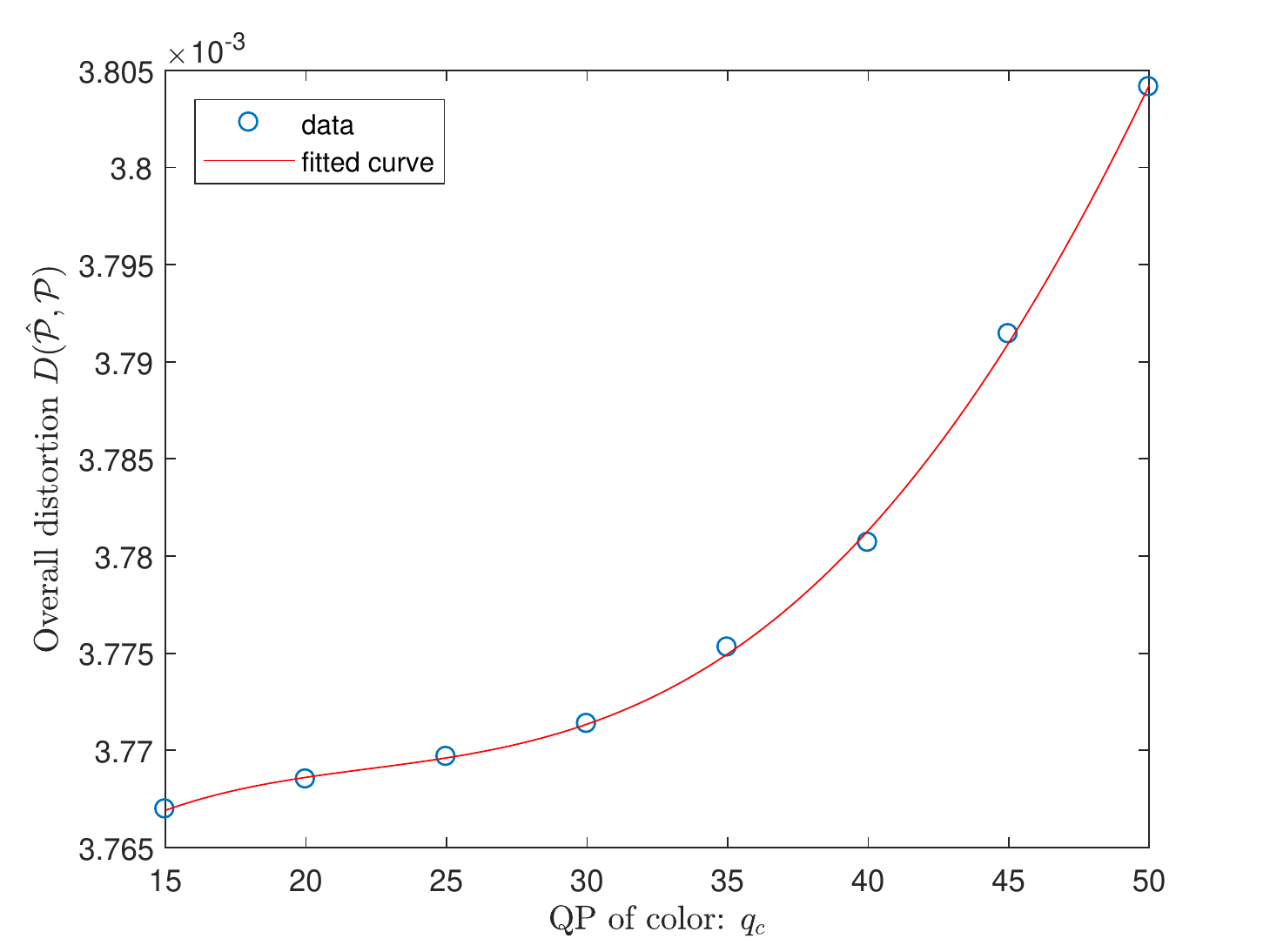}}
	\subfigure[$q_c$ - Soldier]{
		\label{Fig.sub.1}
		\includegraphics[width=0.24\textwidth]{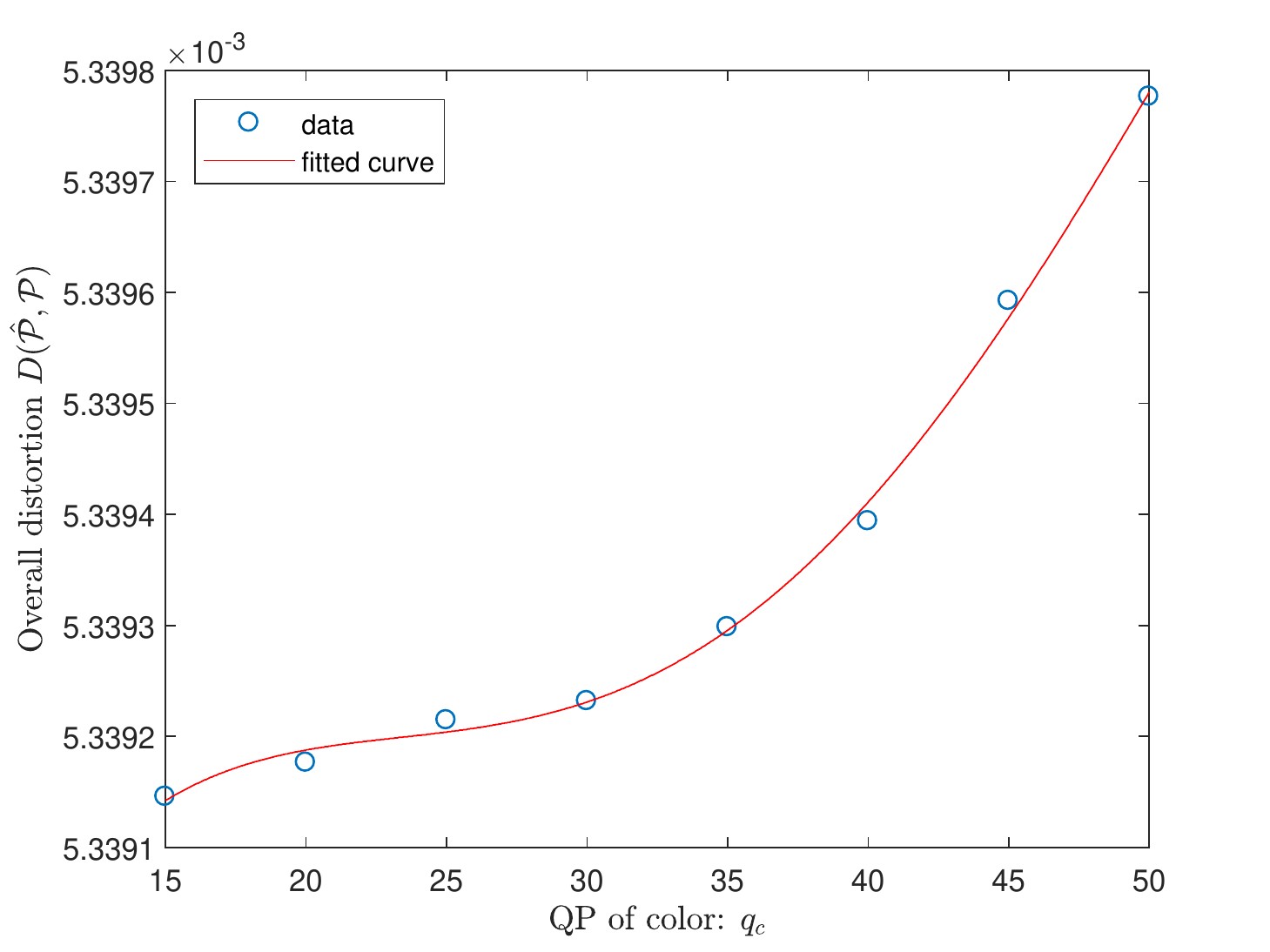}}
	\caption{Polynomial curve fitting for the $D(\hat{\mathcal{P}},{\mathcal{P}})$ and $\rm\bf q$ formed data points, where the fitted functions of $D(\hat{\mathcal{P}},{\mathcal{P}})$ with $q_g$ and $q_c$ are shown in the first and second rows, respectively. (a) $D(\hat{\mathcal{P}},{\mathcal{P}})$ versus $q_g$ for AxeGuy. (b) $D(\hat{\mathcal{P}},{\mathcal{P}})$ versus $q_g$ for Longdress. (c) $D(\hat{\mathcal{P}},{\mathcal{P}})$ versus $q_g$ for Loot. (d) $D(\hat{\mathcal{P}},{\mathcal{P}})$ versus $q_g$ for Soldier. (e) $D(\hat{\mathcal{P}},{\mathcal{P}})$ versus $q_c$ for AxeGuy. (f) $D(\hat{\mathcal{P}},{\mathcal{P}})$ versus $q_c$ for Longdress. (g) $D(\hat{\mathcal{P}},{\mathcal{P}})$ versus $q_c$ for Loot. (h) $D(\hat{\mathcal{P}},{\mathcal{P}})$ versus $q_c$ for Soldier.}
	\label{curve_fitting_distortion_q}
\end{figure*}

\subsection{Color Distortion and Geometry Distortion Unified Model}\label{Unified_model}
To examine the compression distortion of the point clouds, several quality metrics have been proposed in the literature. For geometry distortion, point-to-point and point-to-plane distances are commonly used. In these metrics, the geometry error is usually measured in terms of mean squared error (MSE), root mean squared error (RMS), or Hausdorff distance (Haus) \cite{Hausdorff} by pooling the position errors calculated for individual points. Although geometry distortion is computed using not only point-to-point but also point-to-plane and plane-to-plane correspondences, the color distortion considers only point-to-point correspondences. Since it requires the color information of two points, the color distortion is measured by finding the difference between the corresponding points for each color component separately on a per point basis by calculating the MSE or RMS. In the following, in order to balance these two types of distortion metrics, we develop a unified model. 

Assuming that we have two point clouds, one is the original point cloud $\mathcal{P}$ with the total number of points being $N_\mathcal{P}$, and the other one is compressed point cloud $\hat {\mathcal{P}}$ with a total number of points $ N_{\hat {\mathcal{P}}}$. Denote by $P_i$ one particular point in the point set  $\mathcal{P}$, i.e., $P_i \in \mathcal{P}, 1 \le i \le  N_{ {\mathcal{P}}}$. For the convenience of processing each point, the  geometry position and color attribute   of  $P_i$ are represented as $P_i^g=[x_i,y_i,z_i]$ and $P_i^c=[Y_i,U_i,V_i]$, respectively. Here, we convert the RGB color in the original point clouds to YUV following the ITU-R Rec-BT.709 \cite{BT709}. 
Similarly, we represent one particular point in $\hat {\mathcal{P}}$ as $\hat P_j$, with two attributes $\hat P_j^g=[x_j,y_j,z_j]$ and $\hat P_j^c=[Y_j,U_j,V_j]$, where $ 1 \le j \le  N_{\hat {\mathcal{P}}}$.
Based on these notations, we can have geometry and color distortion expressions. Assume that $\hat {\mathcal{P}}$ is the test point cloud and $ {\mathcal{P}}$ is the reference point cloud. The one-sided point-to-point distance from $\hat {\mathcal{P}}$ to $ {\mathcal{P}}$ with MSE pooling calculated by using L2 norm is shown as follows
\begin{equation}
\label{geometry_distance}
\tilde d_g(\hat {\mathcal{P}}, {\mathcal{P}})=\frac{1}{ N_{\hat {\mathcal{P}}}} \sum_{\forall {\hat P_j \in \hat {\mathcal{P}}}}||\hat P_j^g-P_{i^*}^g||^2_2
\end{equation}
where $P_{i^*}^g$ is the geometry vector of the nearest neighbouring point $i^*$ in ${\mathcal{P}}$ corresponding to $\hat P_j$.  As the one-sided point-to-point distance may lead to significantly underestimated distance values, bidirectional   point-to-point distance is usually adopted as the error metric between two point clouds, which is deduced as follow 
\begin{equation}
 d_g(\hat {\mathcal{P}}, {\mathcal{P}})={ \max}\{\tilde d_g(\hat {\mathcal{P}}, {\mathcal{P}}),\tilde d_g( {\mathcal{P}}, {\hat {\mathcal{P}}})\}
\end{equation}
where $\tilde d_g( {\mathcal{P}}, {\hat {\mathcal{P}}})$ denotes the distance from $ {\mathcal{P}}$ to $\hat {\mathcal{P}}$. When considering the point-to-plane distance between two point sets, the error vector for each test point is multiplied by the normal vector of the associated nearest neighboring point in the reference point cloud, which leads to that \eqref{geometry_distance} is modified to \cite{Dong_Tian}
\begin{equation}
\label{plane_distance}
\tilde d_p(\hat {\mathcal{P}}, {\mathcal{P}})=\frac{1}{ N_{\hat {\mathcal{P}}}} \sum_{\forall {\hat P_j \in \hat {\mathcal{P}}}}||(\hat P_j^g-P_{i^*}^g)N_{i^*}||^2_2
\end{equation}
where $\tilde d_p(\hat {\mathcal{P}}, {\mathcal{P}})$ is the point-to-plane distance  from $\hat {\mathcal{P}}$ to $ {\mathcal{P}}$.  $N_{i^*}$ is the normal vector of $i^*$. Then, the final point-to-plane distance $d_p(\hat {\mathcal{P}}, {\mathcal{P}})$ is derived as the maximum of the two sided distance values, i.e., $\tilde d_p(\hat {\mathcal{P}}, {\mathcal{P}})$ and  $\tilde d_p({\mathcal{P}},\hat {\mathcal{P}})$.

In comparison to computing the Euclidian distance in the form of L2 norm for the geometry distortion, the color distortion of each color component is evaluated by finding the difference between the corresponding points in each color channel. 
For example, the luminance color distortion from $\hat {\mathcal{P}}$ to $ {\mathcal{P}}$  is defined as 
\begin{equation}
\tilde d_{c\_Y}(\hat {\mathcal{P}}, {\mathcal{P}})=\frac{1}{ N_{\hat {\mathcal{P}}}} \sum_{\forall {\hat P_j \in \hat {\mathcal{P}}}}\left(\hat P_j^c(Y)-P_{i^*}^c(Y)\right)^2
\end{equation}
where $P_j^c(Y)$ and $P_{i^*}^c(Y)$ are the luminance of $P_j^c$ and $P_{i^*}^c$, respectively. The bidirectional luminance color distortion $d_{c\_Y}(\hat {\mathcal{P}}, {\mathcal{P}})$ can be then derived using the max operation. In a similar manner, the chrominance distortions $d_{c\_U}(\hat {\mathcal{P}}, {\mathcal{P}})$ and $d_{c\_V}(\hat {\mathcal{P}}, {\mathcal{P}})$ can be determined. 
As human visual system is more sensitive to  luminance distortion than  chrominance distortion, the overall color distortion for point clouds is a weighted average of the  luminance and  chrominance distortions, i.e.,
\begin{equation}
 d_{c}(\hat {\mathcal{P}}, {\mathcal{P}})=\frac{1}{8}(6 \times d_{c\_Y}(\hat {\mathcal{P}}, {\mathcal{P}})+d_{c\_U}(\hat {\mathcal{P}}, {\mathcal{P}})+d_{c\_V}(\hat {\mathcal{P}}, {\mathcal{P}}))
\end{equation}

After having derived geometry and color distortions, i.e., $d_{g}(\hat {\mathcal{P}}, {\mathcal{P}})$ or  $d_{p}(\hat {\mathcal{P}}, {\mathcal{P}})$, and $d_{c}(\hat {\mathcal{P}}, {\mathcal{P}})$, we attempt to derive a unified distortion model.  { As geometry and color distortions are measured at different scales, we derive the covariance matrix between geometry and color components of point clouds, which gives the covariance between these two elements, and thus can be used to balance the geometry and color distortions.}
We denote the weighted averages of the geometry component values and the color componet values for the point $P_i$ in $\mathcal{P}$ by $\overline { P_i^g}$ and $\overline { P_i^c}$, respectively. In accordance with the calculations for the geometry and overall color distortions, they are calculated as follows
\begin{equation}
\begin{aligned}
&\overline { P_i^g}=\frac{1}{3}(x_i+y_i+z_i)\\
&\overline { P_i^c}=\frac{1}{8}(6Y_i+U_i+V_i)
\end{aligned}
\end{equation}

We then concatenate  $\overline { P_i^g}$ and $\overline { P_i^c}$ into a two-element attribute vector $\overline {P_i}$ for each point, \emph{i.e.}, $\overline {P_i}=[\overline { P_i^g},\overline { P_i^c}],\forall i \in N_{\mathcal{P}}$. With $\overline {P_i}$, the covariance matrix $S_{\mathcal{P}}$ between geometry and color for ${\mathcal{P}}$ can be written as 
\begin{equation}
\label{covariance_matrix_original_points}
S_{\mathcal{P}}=\frac{1}{N_{\mathcal{P}}} \sum_{i=1}^{ N_{\mathcal{P}}}(\overline {P_i}-\overline{P})^T(\overline {P_i}-\overline{P})
\end{equation}
where $\overline{P}=[\overline { P^g},\overline { P^c}]$, in which 	$\overline { P^g}$ and $\overline { P^c}$ are the averages of 	$\overline { P_i^g}$ and $\overline { P_i^c}$ of all the points in the $\mathcal{P}$, respectively, i.e.,
\begin{equation}
\begin{aligned}
\overline { P^g}=\frac{1}{N_{\mathcal{P}}} \sum_{i=1}^{ N_{\mathcal{P}}}\overline { P_i^g}\\
\overline { P^c}=\frac{1}{N_{\mathcal{P}}} \sum_{i=1}^{ N_{\mathcal{P}}}\overline { P_i^c}
\end{aligned}
\end{equation}

Similar to \eqref{covariance_matrix_original_points}, we can have the covariance matrix for the compressed point clouds ${\mathcal{\hat P}}$, i.e., $S_{\mathcal{\hat P}}$. The final covariance matrix can thus be obtained by pooling the covariance matrices of the original and compressed point clouds, which is given as follows
\begin{equation}
S(\hat{\mathcal{P}},{\mathcal{P}})=\frac{N_{\mathcal{P}}}{N_{\mathcal{P}}+N_{\hat{\mathcal{P}}}}{S_{\mathcal{P}}}+\frac{N_{\hat {\mathcal{P}}}}{N_{\mathcal{P}}+N_{\hat{\mathcal{P}}}}{S_{\hat{\mathcal{P}}}}
\end{equation}

We define $D_{gc}(\hat{\mathcal{P}},{\mathcal{P}})$ as the distortion vector measured for two point clouds, which is formed by concatenating the measured geometry and color distortions, i.e., $D_{gc}(\hat{\mathcal{P}},{\mathcal{P}})=[d_{g}(\hat {\mathcal{P}}, {\mathcal{P}}),d_{c}(\hat {\mathcal{P}}, {\mathcal{P}})]$ or $D_{gc}(\hat{\mathcal{P}},{\mathcal{P}})=[d_{p}(\hat {\mathcal{P}}, {\mathcal{P}}),d_{c}(\hat {\mathcal{P}}, {\mathcal{P}})]$. 
In this work, we use the point-to-point distance, i.e., $d_{g}(\hat {\mathcal{P}}, {\mathcal{P}})$, as the input of the geometry distortion to the overall distortion due to its better correlation with the subjective scores. The statistical correlations of objective quality metrics with the subjective scores will be analysed in Section \ref{Experiment_A}. 
By using the covariance matrix $S$, we can unify the geometry and color distortion as follows
\begin{equation}\label{Unified_distortion_model}
D(\hat{\mathcal{P}},{\mathcal{P}})=\sqrt{(D_{gc}(\hat{\mathcal{P}},{\mathcal{P}}))(S(\hat{\mathcal{P}},{\mathcal{P}}))^{-1} (D_{gc}(\hat{\mathcal{P}},{\mathcal{P}}))^T}
\end{equation}
where $(S(\hat{\mathcal{P}},{\mathcal{P}}))^{-1}$ is the inverse of the matrix $S(\hat{\mathcal{P}},{\mathcal{P}})$. $D(\hat{\mathcal{P}},{\mathcal{P}})$ is the overall distortion metric for point clouds, which can be employed for bit rate constrained compression and optimization.

\begin{figure*}[!htbp]
	\hspace{-1mm}\subfigure[$q_g$ - AxeGuy]{
		\label{Fig.sub.1}
		\includegraphics[width=0.24\textwidth]{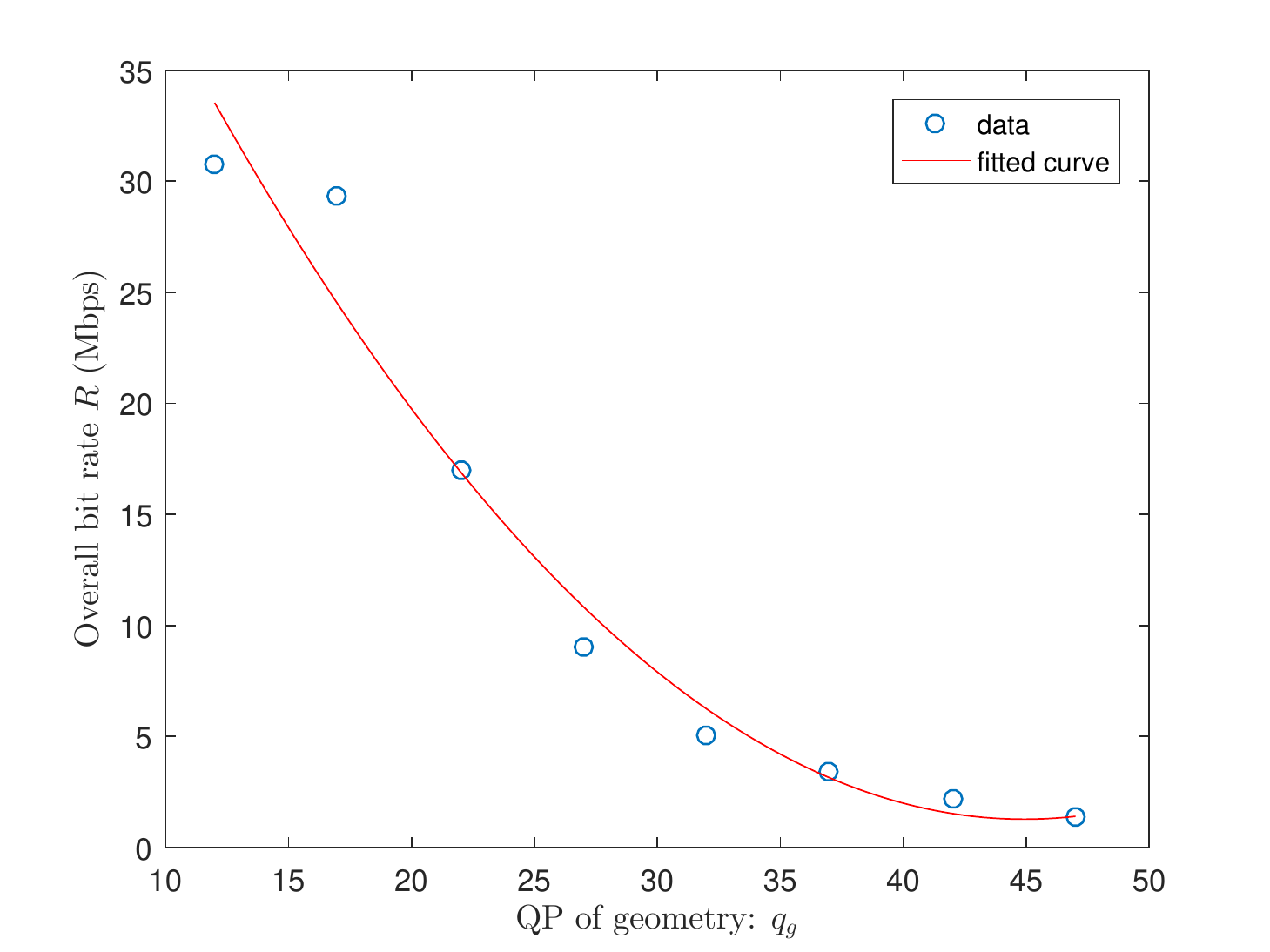}}
	\subfigure[$q_g$ - Longdress]{
		\label{Fig.sub.1}
		\includegraphics[width=0.24\textwidth]{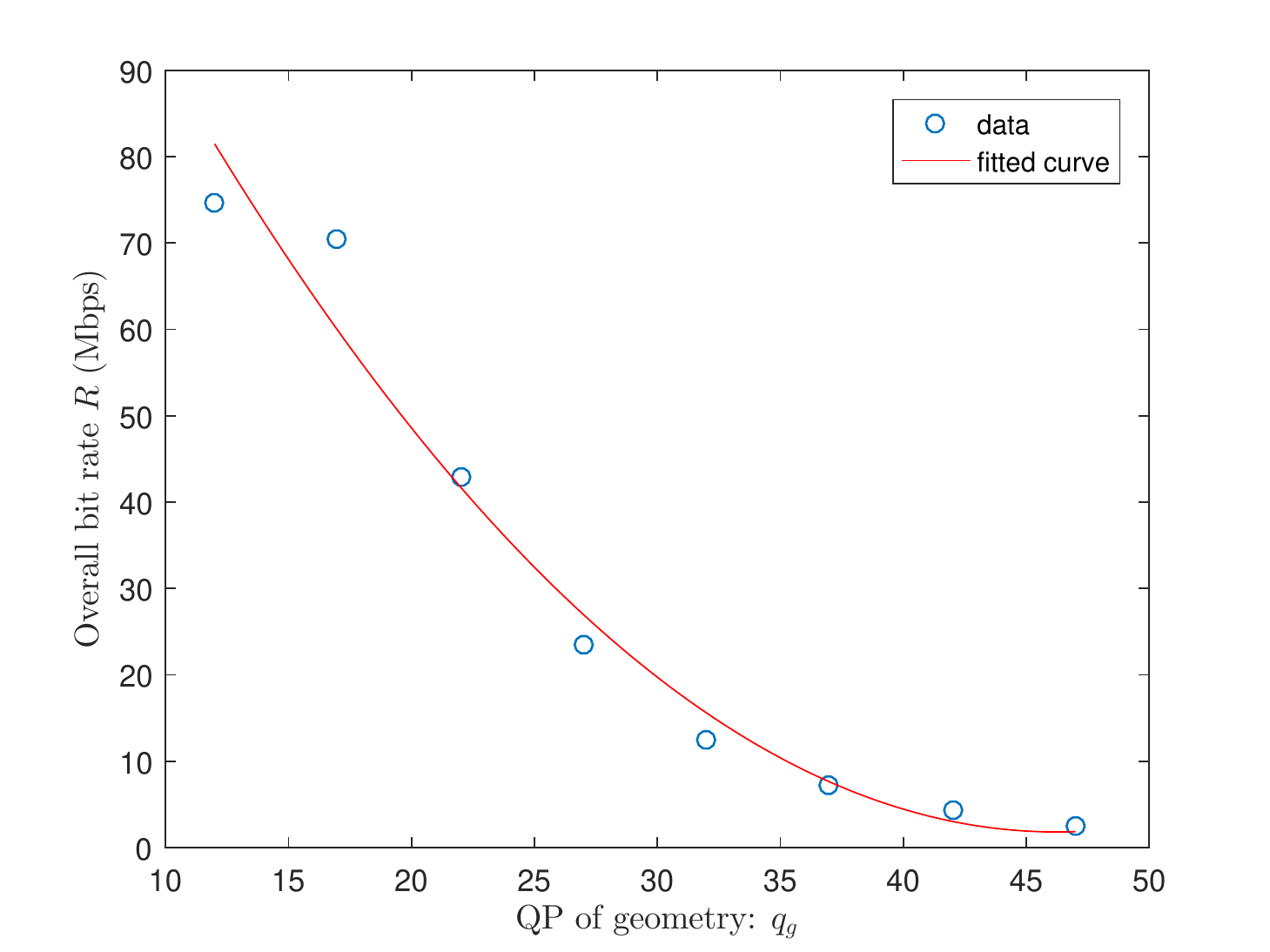}}
	\subfigure[$q_g$ - Loot]{
		\label{Fig.sub.1}
		\includegraphics[width=0.24\textwidth]{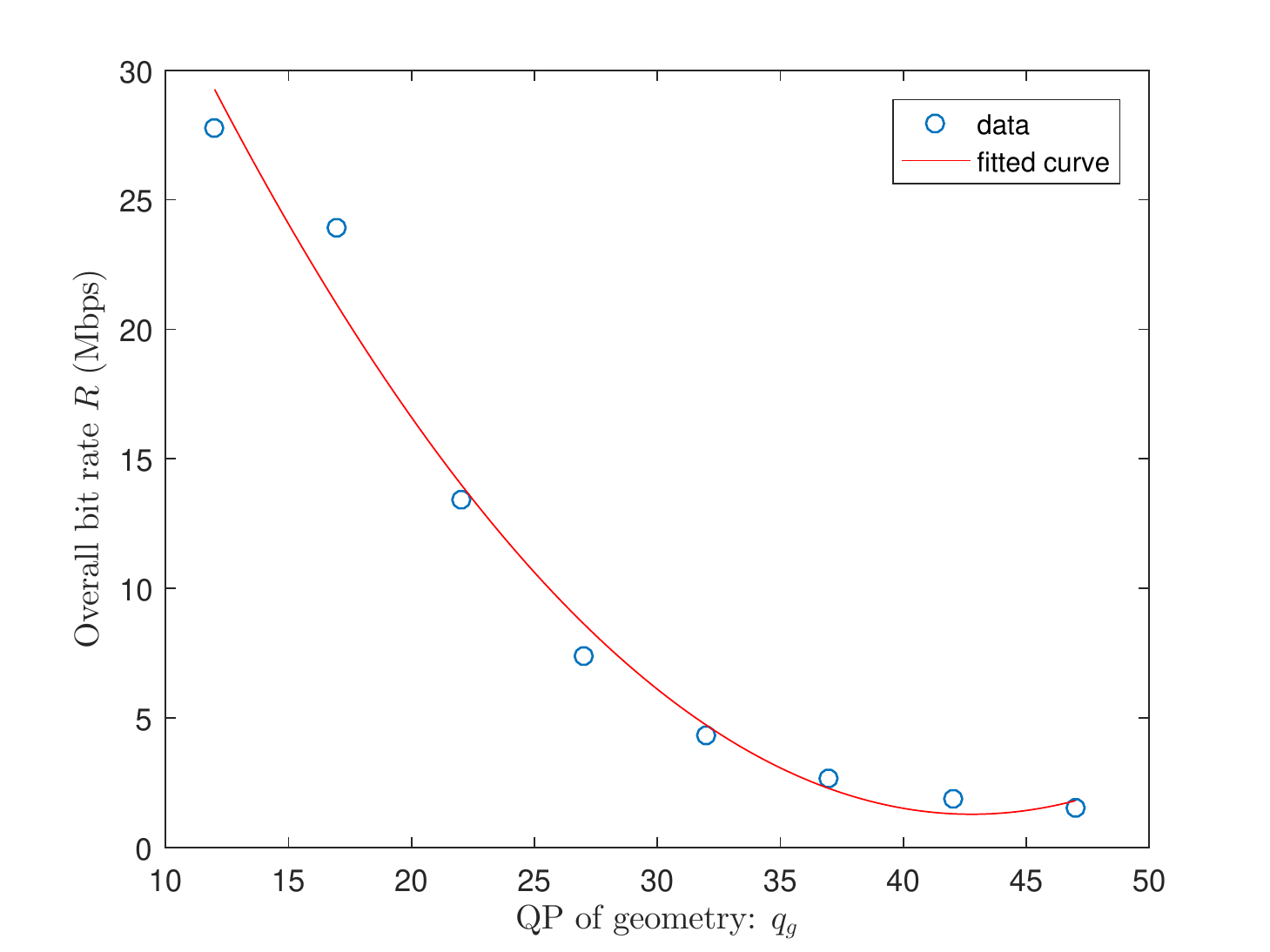}}
	\subfigure[$q_g$ - Soldier]{
		\label{Fig.sub.1}
		\includegraphics[width=0.24\textwidth]{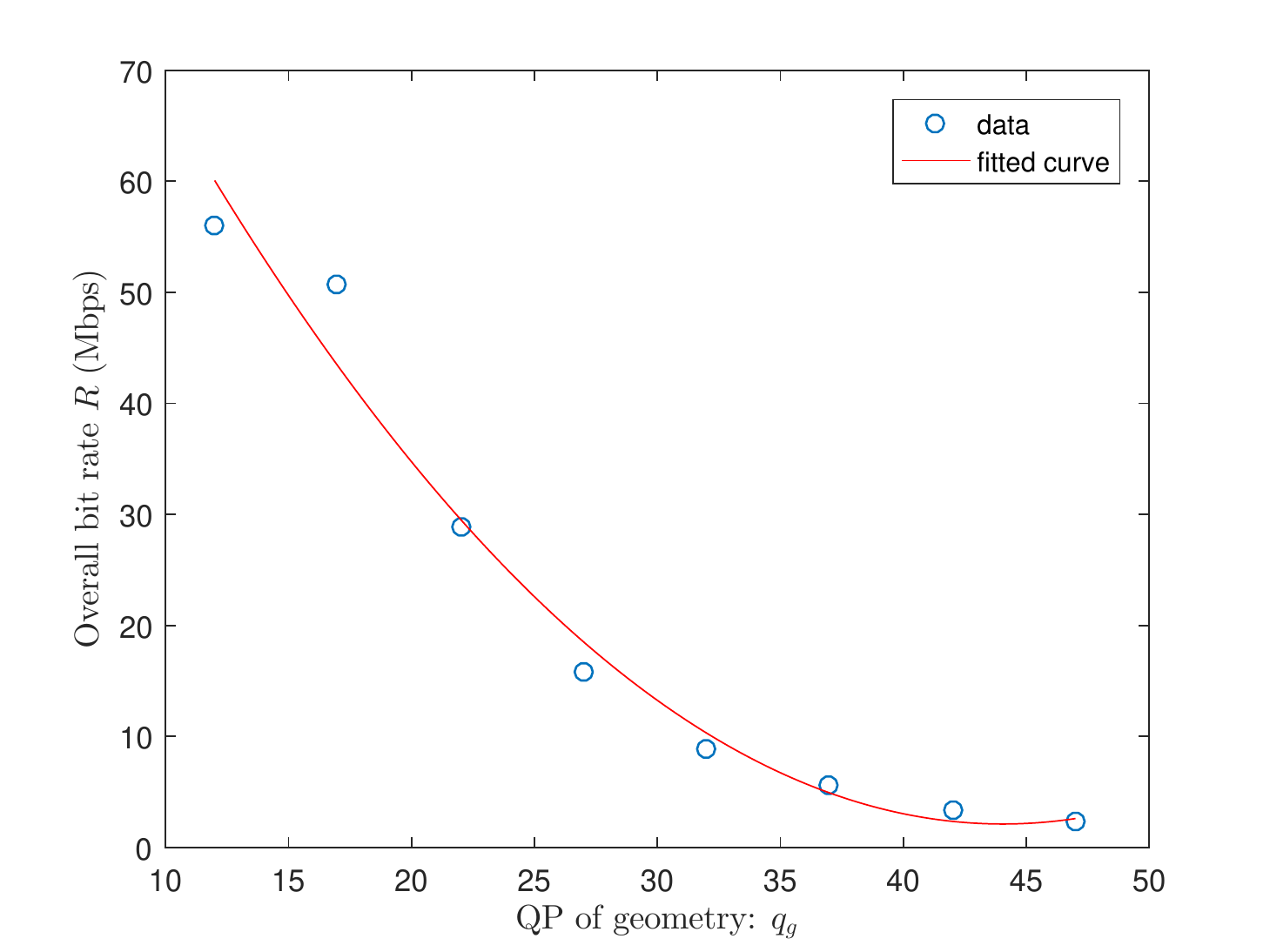}}
	
	\hspace{-1mm}\subfigure[$q_c$ - AxeGuy]{
		\label{Fig.sub.1}
		\includegraphics[width=0.24\textwidth]{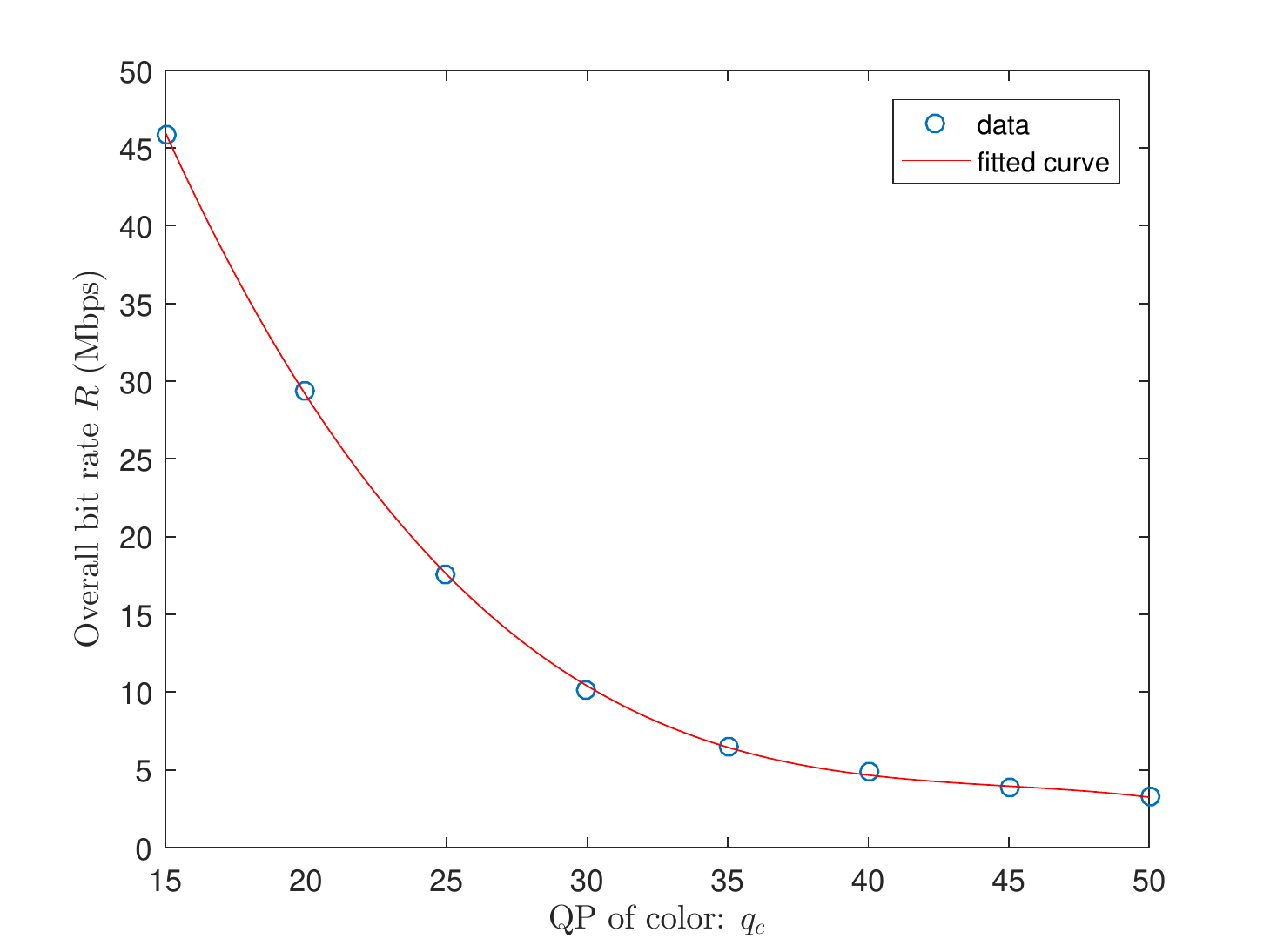}}
	\subfigure[$q_c$ - Longdress]{
		\label{Fig.sub.1}
		\includegraphics[width=0.24\textwidth]{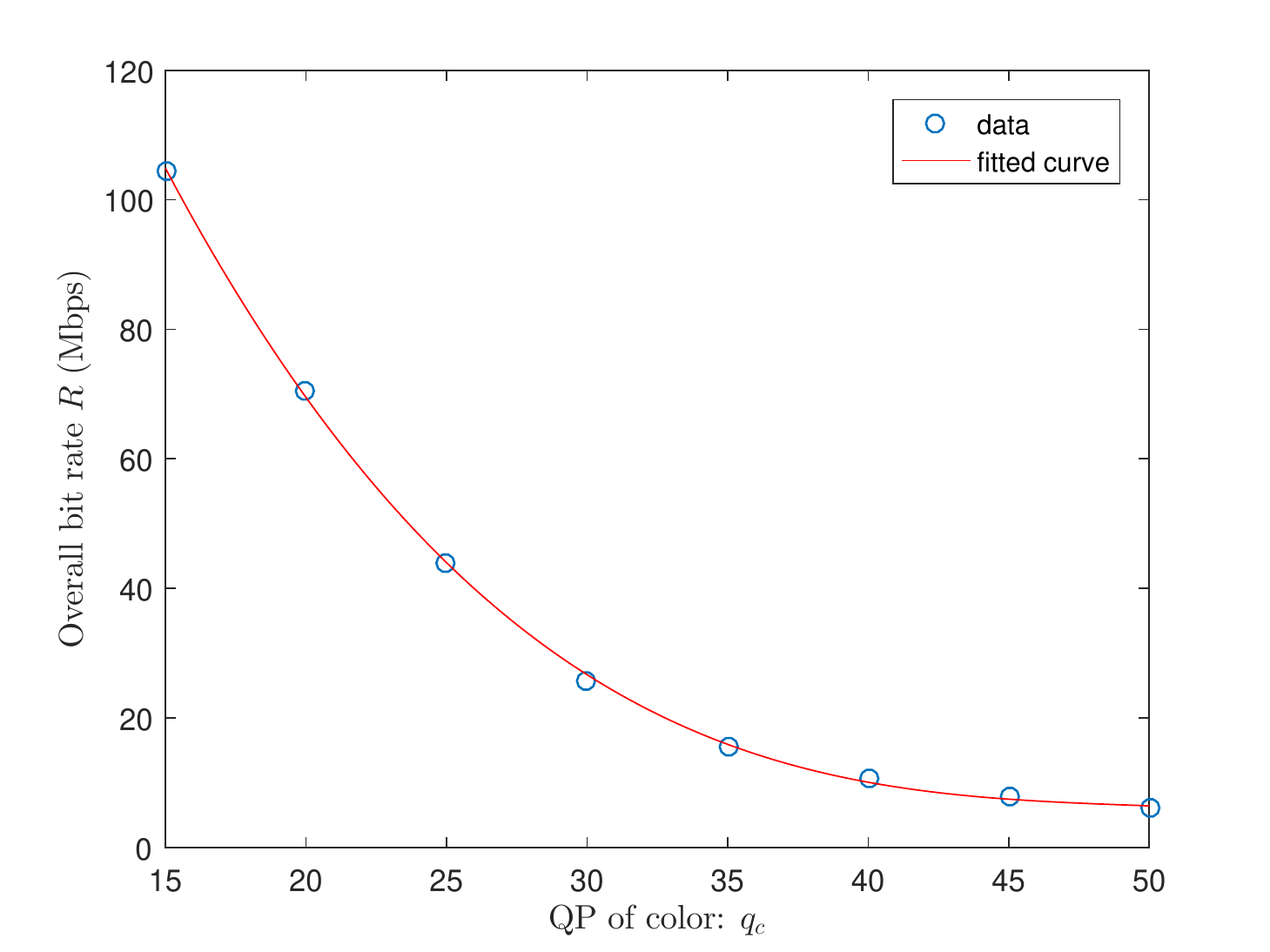}}
	\subfigure[$q_c$ - Loot]{
		\label{Fig.sub.1}
		\includegraphics[width=0.24\textwidth]{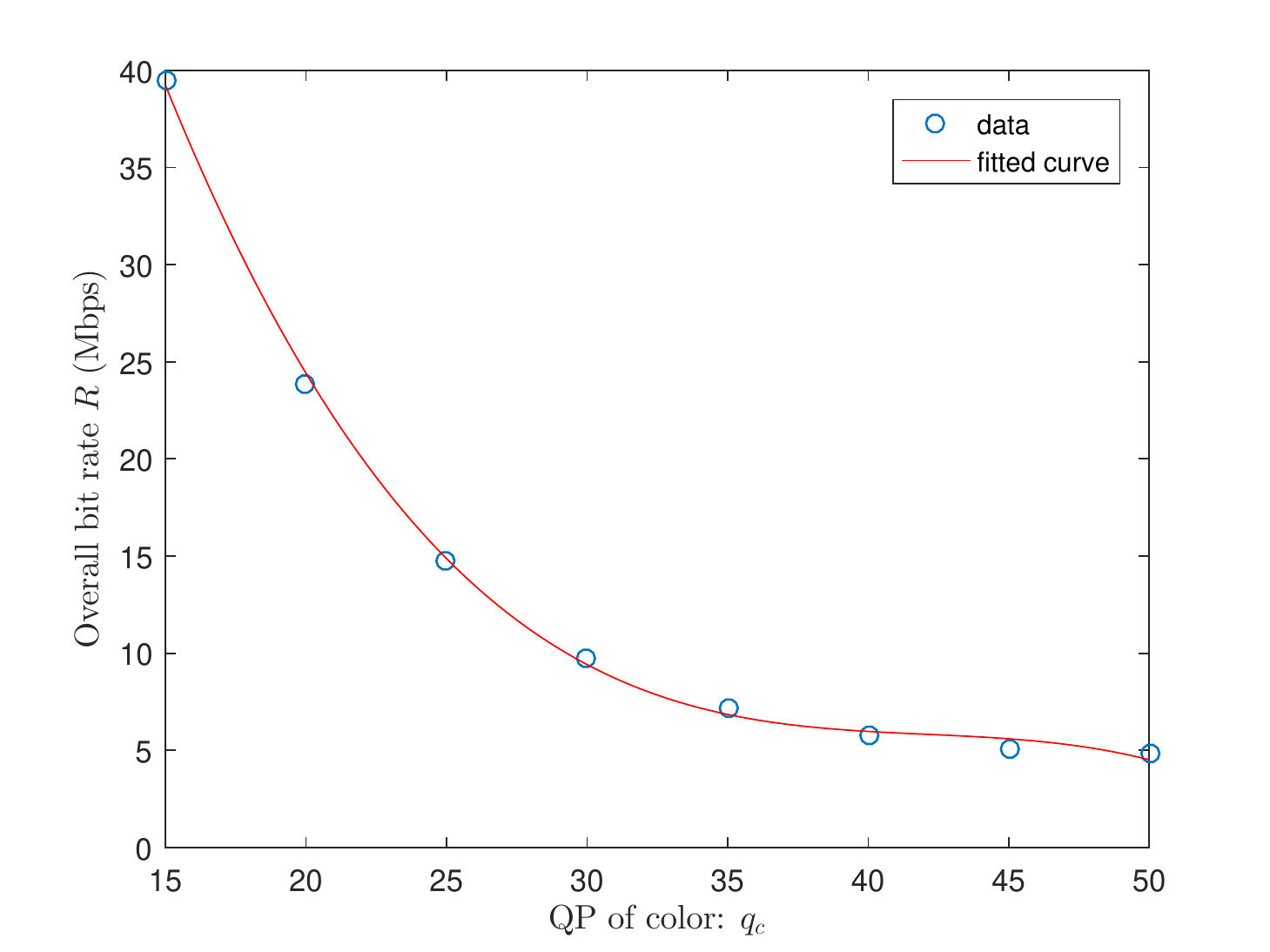}}
	\subfigure[$q_c$ - Soldier]{
		\label{Fig.sub.1}
		\includegraphics[width=0.24\textwidth]{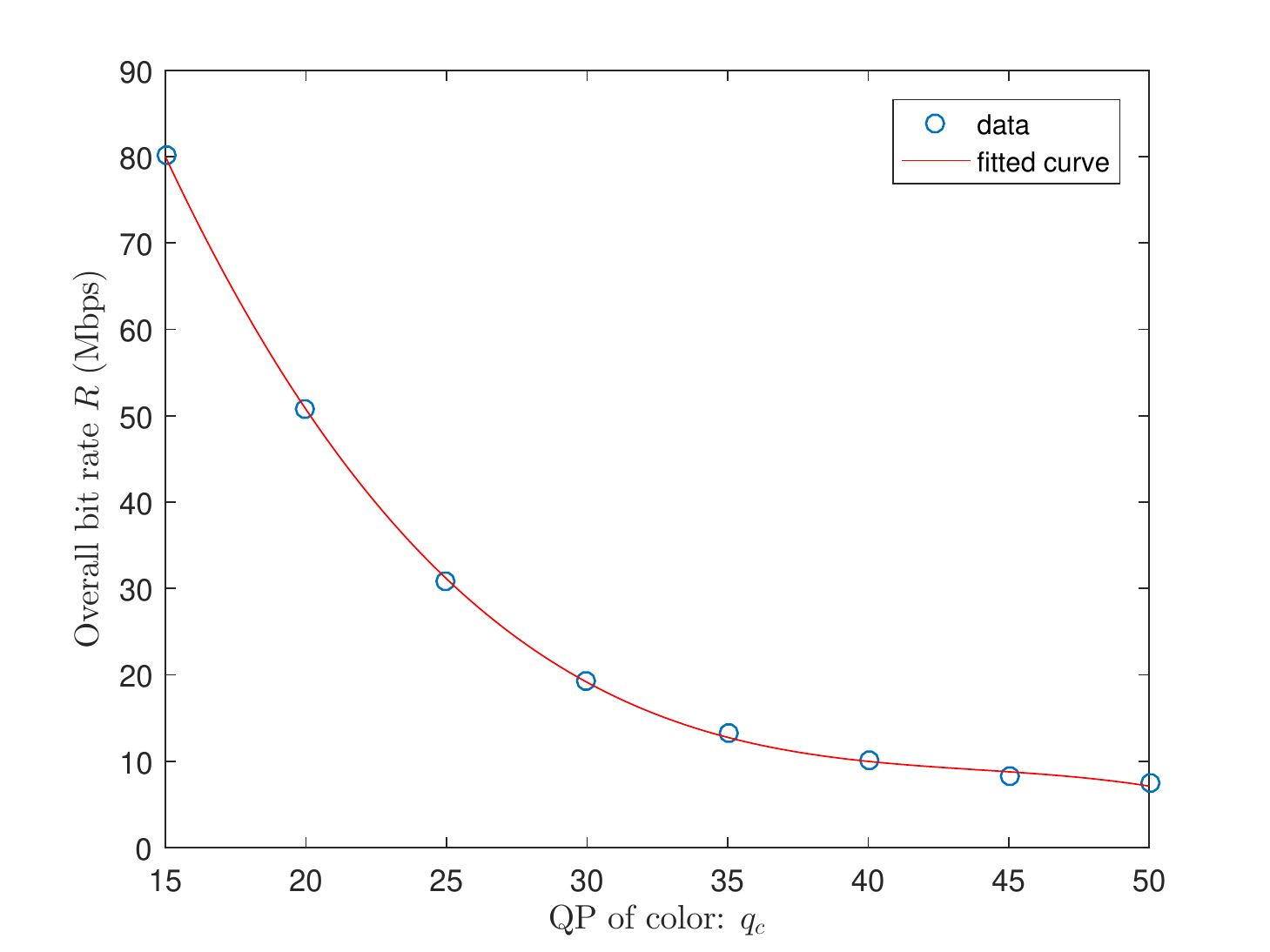}}
	\caption{Polynomial curve fitting for the bit rate $R$ and $\rm\bf q$ formed data points, where the fitted functions of $R$ with $q_g$ and $q_c$ are shown in the first and second rows, respectively. (a) $R$ versus $q_g$ for AxeGuy. (b) $R$ versus $q_g$ for Longdress. (c) $R$ versus $q_g$ for Loot. (d) $R$ versus $q_g$ for Soldier. (e) $R$ versus $q_c$ for AxeGuy. (f) $R$ versus $q_c$ for Longdress. (g) $R$ versus $q_c$ for Loot. (h) $R$ versus $q_c$ for Soldier.}
	\label{curve_fitting_bit_q}
\end{figure*}

\subsection{Relationships Between  Distortion and Bit Rate with QP}
In this subsection, we investigate the relationship between $D(\hat{\mathcal{P}},{\mathcal{P}})$ and $R$ with $\rm\bf q$. To this end, we compress the point clouds with different $\rm\bf q$, and then calculate the corresponding $D(\hat{\mathcal{P}},{\mathcal{P}})$. We fit  polynomial functions with these $D(\hat{\mathcal{P}},{\mathcal{P}})$ and $\rm\bf q$ formed data points. As $\rm\bf q$ involves two QPs, we fit the relationships of the distortion with $q_g$ and $q_c$ one-by-one. That is, when evaluating the relationship of $D(\hat{\mathcal{P}},{\mathcal{P}})$ and $q_g$, we fix $q_c$, and vary $q_g$, and vice versa. The polynomial curve fitting for the data  $D(\hat{\mathcal{P}},{\mathcal{P}})$ and $\rm\bf q$ are illustrated in Fig. \ref{curve_fitting_distortion_q}, where a number of point cloud sequences are used for test, including the ``AxeGuy'' sequence from VSENSE VVDB2 database \cite{Electronic_imaging,zerman2020textured, VVDB2}, and  the common MPEG test sequences, ``Loot'', ``Longdress'', and ``Soldier'' \cite{8i_sequence}. By using the polynomial fitting, we obtained two four-degree polynomials for representing the functions of $D(\hat{\mathcal{P}},{\mathcal{P}})$ with $q_c$ and $q_c$, respectively, which are shown as follows
\begin{equation}
\label{fitted_distortion_fun}
\begin{aligned}
&D(q_g)=a_4{q_g^4}+a_3{q_g^3}+a_2{q_g^2}+a_1{q_g}+a_0\\
&D(q_c)=b_4{q_c^4}+b_3{q_c^3}+b_2{q_c^2}+b_1{q_c}+b_0
\end{aligned}
\end{equation}

By a slight abuse of notation, in \eqref{fitted_distortion_fun}, we use $D(q_g)$ and $D(q_c)$ to represent the functions of $D(\hat{\mathcal{P}},{\mathcal{P}})$ with the arguments of $q_g$ and $q_c$, respectively. $a_4,\cdots,a_0$ and $b_4,\cdots,b_0$ are the  coefficients of the associated fitted functions, which are point cloud sequence dependent. 

Similarly, we perform polynomial fitting for the collected bit rate data, which are shown in Fig. \ref{curve_fitting_bit_q}. From this figure, we obtain a 2-degree polynomial for the relationship of $R$ and $q_g$, represented as $R(q_g)$, and a 3-degree polynomial for  $R$ and $q_c$, i.e., $R(q_c)$. The specific expressions of $R(q_g)$ and $R(q_c)$ are written as follows
\begin{equation}
\label{fitted_bit_fun}
\begin{aligned}
&R(q_g)=c_2{q_g^2}+c_1{q_g}+c_0\\
&R(q_c)=d_3{q_c^3}+d_2{q_c^2}+d_1{q_c}+d_0
\end{aligned}
\end{equation}
where $c_2\cdots c_0$ and $d_3\cdots d_0$ are the coefficients of these functions. 

By combining the $D(q_g)$ with $D(q_c)$  in \eqref{fitted_distortion_fun} and $R(q_g)$ with $R(q_c)$ in \eqref{fitted_bit_fun}, we can have the overall distortion $D$ and $\rm\bf q$ relationship, i.e., $D(q_g,q_c)$, and, the bit rate $R$ and $\rm\bf q$ relationship, i.e., $R(q_g,q_c)$, for point cloud compression, respectively, which are shown below
\begin{equation}
\label{fitted_distortion_overall}
\begin{aligned}
D(q_g,q_c)=&a_4{q_g^4}+b_4{q_c^4}+a_3{q_g^3}+b_3{q_c^3}\\
&+a_2{q_g^2}+b_2{q_c^2}+a_1{q_g}+b_1{q_c}+a_0+b_0
\end{aligned}
\end{equation}

\begin{equation}
\label{fitted_distortion_overall}
\begin{aligned}
R(q_g,q_c)=&d_3{q_c^3}+d_2{q_c^2}+c_2{q_g^2}\\
&+d_1{q_c}+c_1{q_g}+d_0+c_0
\end{aligned}
\end{equation}

{To determine the parameters in (16) and (17), we need to pre-encode the test point cloud. More specifically, to determine the parameters associated with $q_g$ in (16), we employ five pairs of geometry and color QPs to encode, i.e., $(33,35), (30,35), (26,35), (20, 35), (15, 35)$, in which the $q_g$ is varied given a specific $q_c$. We can thus obtain five polynomial equations using the $D(q_g)$ function in (14), and finally solve them to get  $a_4,\cdots, a_0$. Similarly, to determine the remaining five parameters in (16), we also need to encode the point cloud using five QP pairs, but with $q_c$ being varied given a specific $q_g$. For this purpose, the QP pairs 
we adopt are $(30,38), (30,35), (30,31), (30,26), (30,20)$. As $(30,35)$  has been already used once, we only need to perform pre-encoding using the rest four QP pairs. We can then  establish another five  polynomial equations using the $D(q_c)$ function in (14), and derive the parameters $b_4,\cdots, b_0$. With these  pre-encoding results by nine pairs of QPs, we can also build the polynomial equations for $R(q_g)$ and $R(q_c)$ in (15), and finally derive the rate model parameters $c_2,\cdots, c_0$ and $d_3,\cdots, d_0$.}

\section{Rate Distortion Optimization for Point Clouds}\label{RDO_PCC}
In this section, we perform rate distortion optimization for point cloud compression using the distortion and rate models derived in the previous section. As in \eqref{constrained_compression_formula}, the bit rate constrained compression  for point cloud can be formulated as follows
\begin{equation}
\begin{aligned}
& [{q_g^*,q_c^*}]=\operatorname*{arg\,min}_{[{q_g,q_c}]} D({q_g,q_c})\\
&  \text{s. t.} \;
R({{q_g,q_c}}) \leq \widehat R \\
\end{aligned}
\label{constrained_point_cloud_optimization}
\end{equation}

To solve \eqref{constrained_point_cloud_optimization}, in theory, we can use the Lagrange multiplier method introduced in Section I to relax the constraint and then 
derive the optimal solution using the necessary optimality conditions. However, as the  distortion model $D({q_g,q_c})$ and rate model $R({{q_g,q_c}}) $ for point clouds are high degree polynomials, the resulting derivative  functions of  $D({q_g,q_c})$ (or $R({{q_g,q_c}}) $) with respect to $\rm\bf q$ also have high degree, which induces that finding the roots of the gradients of the  Lagrangian cost function is not trivial. In this work, we use the augmented Lagrangian method \cite{Augumented_larangian} to solve the constrained problem in \eqref{constrained_point_cloud_optimization}, in which we introduce an additional penalty term into the Lagrangian of the constrained problem. 

The unconstrained objective function for  \eqref{constrained_point_cloud_optimization} using the augmented Lagrangian method can be formulated as follows
\begin{equation}
\begin{aligned}
J({q_g},{q_c},\lambda,\rho)=&D({q_g,q_c})+\lambda (R({{q_g,q_c}})-\widehat R )\\
&+\frac{\rho}{2}(R({{q_g,q_c}})-\widehat R )^2
\end{aligned}
\label{eq_Augmented_Lagrangian}
\end{equation}
where $\rho$ is the penalty coefficient. In the following, we perform a sequential of minimizations of  the augmented Lagrangian functions to approximate the optimal solution. Firstly, we initialize $\lambda$ and $\rho$ with two constants, respectively. Then, at the ${(k+1)}^{th}$ iteration, the values of $\lambda$ and $\rho$, denoted by $\lambda^{k+1}$ and $\rho^{k+1}$, are updated as follows
\begin{equation}
\rho^{k+1}=\alpha \rho^{k}
\label{update_rho}
\end{equation}
\begin{equation}
\lambda^{k+1}= \lambda^{k}+\rho^{k}(R({{q_g^{k+1},q_c^{k+1}}})-\widehat R )
\label{update_lambda}
\end{equation}

In \eqref{update_rho}, for better trade-off between convergence rate and accuracy, we choose the increasing rate $\alpha$ equal to $1.5$. In \eqref{update_lambda}, $q_g^{k+1}$ and $q_c^{k+1}$ are the solution to the unconstrained problem at the $k^{th}$ step, which are represented as follows
\begin{equation}
[q_g^{k+1},q_c^{k+1}]=\argmin_{[{q_g},{q_c}]} J({q_g},{q_c},\lambda^k,\rho^k)
\label{equation_each_iteration}
\end{equation}

We perform minimization of the augmented Lagrangian iteratively using \eqref{update_rho} and \eqref{update_lambda} until the solution to the  augmented Lagrangian converges. In this work, 20 $\sim$ 30 iterations are usually observed to achieve convergences.
{We identify the convergence by checking whether the change in the objective function tends to  approach a predefined minimum value 0.3.}
{ Note that, as $\rho$ is eventually large, quadratic penalty makes the objective function in \eqref{eq_Augmented_Lagrangian}  strongly convex \cite{Augumented_larangian}.} 

For given $\lambda^k$ and $\rho^k$, we use the gradient descent algorithm to find the solution to \eqref{equation_each_iteration}. The gradients of the augmented Lagrangian function with respect to ${q_g}$ and ${q_c}$ are written as follows
\begin{equation}
\begin{aligned}
g_{q_g}(q_g,q_c)=\nabla_{{q_g}} J({q_g},{q_c},\lambda^k,\rho^k)\\
g_{q_c}(q_g,q_c)=\nabla_{{q_c}} J({q_g},{q_c},\lambda^k,\rho^k)
\end{aligned}
\end{equation}

The iteration index within the gradient descent algorithm is denoted by $w$, and the estimates of $q_g$ and $q_c$ at iteration $w$ are denoted by $q_g^w$ and $q_c^w$, respectively. At the ${(w+1)}^{th}$ step, $q_g^{w+1}$ and $q_c^{w+1}$ can be derived as follows
\begin{equation}
\begin{aligned}
q_g^{w+1}=q_g^{w}-\gamma g_{q_g}(q_g^w,q_c^w)\\
q_c^{w+1}=q_c^{w}-\gamma g_{q_c}(q_g^w,q_c^w)
\end{aligned}
\label{iteration_gradient}
\end{equation}
where $\gamma$ is the step size in the gradient descent method. 
{The iteration of \eqref{iteration_gradient} continues until the decrement of the objective function $J({q_g},{q_c},\lambda^k,\rho^k)$ falls below a small threshold, i.e., 0.3, as shown in the Algorithm~\ref{alg:RDO}. 
Note that, to avoid ill-conditioning at the first minimization, we set $\rho^0$ to 50, and $\lambda^0$ is set to zero for simplicity as done in \cite{Augumented_larangian}. In the submodule of gradient descent, we set $\gamma$ to the commonly used value 0.001. In addition, we consider the QP range as $[2,\cdots,51]$ in this work, and, $q_g^0$ and $q_c^0$ are thus both initialized to 51.}

\begin{algorithm}
	\caption{Rate-distortion optimization for point cloud compression}
	\label{alg:RDO}
	\begin{algorithmic}[1]
		
		\State Set $k=0$, and choose $\lambda^0$ and $\rho^0>0$
		\Repeat
		\State $[q_g^{k+1},q_c^{k+1}]=\argmin_{[{q_g},{q_c}]} J({q_g},{q_c},\lambda^k,\rho^k)$
		\State $\rho^{k+1}=\alpha \rho^{k}$
		\State $\lambda^{k+1}= \lambda^{k}+\rho^{k}(R({{q_g^{k+1},q_c^{k+1}}})-\widehat R )$
		\State $k \leftarrow k+1$
		\Until{The  augmented Lagrangian cost converges.}
		
		\Procedure{$\argmin_{[{q_g},{q_c}]} $}{$J({q_g},{q_c},\lambda^k,\rho^k)$} 
		\State Set $w=0$, and choose $\gamma>0$	
		\State Initialize  $q_g^0$ and $q_c^0$
		\Repeat
		\State $q_g^{w+1}=q_g^{w}-\gamma g_{q_g}(q_g^w,q_c^w)$ 
		\State $q_c^{w+1}=q_c^{w}-\gamma g_{q_c}(q_g^w,q_c^w)$
		\State $w \leftarrow w+1$
		\Until{Stopping criterion is satisfied.}
		\EndProcedure
		
	\end{algorithmic}
\end{algorithm}

\section{Experimental Results and Discussions}\label{Experiments}
We validate the proposed bit-rate constrained rate-distortion optimization algorithm on the MPEG V-PCC reference software, also known as TMC2 \cite{Video-Test-model, TMC2}.
The coding parameters are set according to the point cloud compression common test conditions \cite{G_PCC_CTC}. 
{More specifically, colorTransform is set to zero, minPointCountPerCCPatchSegmentation is 16, and surfaceThickness is set to 1. The maximum distances for a point to be ignored
during raw points detection and selection  are 9 and 1, respectively. The occupancyResolution is set to 16. During encoding of converted geometry and texture videos, HM-16.20 reference software is used, where the Profile of main10 is selected and the GOP size is 16.}
The point cloud sequence selected here for test includes AxeGuy, Longdress, Loot, Soldier, Ricardo10 \cite{Upperbody}, Phil9 \cite{Upperbody}. The point count for each frame of them is about 400K, 850K, 800K, 1000K, 950K, 350K, respectively. 
Four target bit rates are chosen for each test point cloud sequence. For each target bit rate, we use the proposed algorithm to determine the optimal QPs for color coding and geometry coding. We then use the determined QPs as the input to V-PCC for encoding the point clouds. The unified distortion model proposed in Section~\ref{Unified_model} is used to evaluate the quality of the compressed point clouds. Analogously to quality evaluation in color image/video, in order to better understand quality degradation between point clouds, we convert the unified distortion to PSNR. The PSNR used for measuring the quality of the point clouds is dubbed PC-PSNR, which is derived as follows
\begin{equation}
PC\hspace{-1mm}-\hspace{-1mm}PSNR=10\log_{10}\left({\frac{{MAX}^2}{D(\hat{\mathcal{P}},{\mathcal{P}})}}\right)
\end{equation}
where $D(\hat{\mathcal{P}},{\mathcal{P}})$ is the unified distortion for point clouds calculated in \eqref{Unified_distortion_model}. $MAX$ is the peak value to be determined. As the unified distortion has been normalized using the covariance matrix of geometry component and color component of point clouds, we set $MAX$ to 2 empirically in this work. 

\begin{table}[tbp]
	\caption{Correlation comparison of objective quality metrics with the subjective quality scores from the VVDB2 database \cite{zerman2020textured, VVDB2}.}
	\label{Correlation-Coefficient}
	\begin{tabular}{|l|l|l|l|l|}
		\hline
		Metrics               & PCC   & SROCC & RMS  & OR    \\ \hline
		$p2point_{RMS}$       & 0.915 & 0.805 & 7.85  & 0.046 \\ \hline
		$p2point_{Haus}$      & 0.089 & 0.056 & 20.25 & 0.406 \\ \hline
		$p2plane_{RMS}$       & 0.665 & 0.738 & 15.18 & 0.227 \\ \hline
		$p2plane_{Haus}$      & 0.087 & 0.055 & 20.25 & 0.406 \\ \hline
		$pl2plane_{mean}$     & 0.287 & 0.118 & 19.47 & 0.359 \\ \hline
		$pl2plane_{RMS}$      & 0.287 & 0.112 & 19.47 & 0.359 \\ \hline
		$pl2plane_{MSE}$      & 0.287 & 0.112 & 19.47 & 0.359 \\ \hline
		$MSE_Y$               & 0.903 & 0.874 & 8.72  & 0.070 \\ \hline
		$MSE_U$               & 0.555 & 0.644 & 16.92 & 0.336 \\ \hline
		$MSE_V$               & 0.567 & 0.583 & 16.75 & 0.367 \\ \hline
		$PSNR-p2point_{Haus}$ & 0.257 & 0.280 & 19.64 & 0.391 \\ \hline
		$PSNR-p2plane_{RMS}$  & 0.641 & 0.702 & 15.60 & 0.234 \\ \hline
		$PSNR-p2plane_{Haus}$ & 0.443 & 0.500 & 18.22 & 0.273 \\ \hline
		$PSNR_Y$              & 0.912 & 0.878 & 8.35  & 0.094 \\ \hline
		$PSNR_U$              & 0.708 & 0.658 & 14.35 & 0.266 \\ \hline
		$PSNR_V$              & 0.659 & 0.593 & 15.29 & 0.289 \\ \hline
		$PSNR_{YUV}$          & 0.887 & 0.849 & 9.39  & 0.141 \\ \hline
		$PCQM$                & 0.90 & 0.83 & 7.71  & 0.041 \\ \hline
		PC-PSNR               & 0.928 & 0.893 & 7.59  & 0.031 \\ \hline
	\end{tabular}
\end{table}
{
\subsection{Validation of the Proposed Distortion Unified Model}\label{Experiment_A}

In this subsection, we verify the performance of the proposed  unified distortion model, i.e., the proposed objective metric PC-PSNR. For this purpose, we use the VSENSE VVDB2 database \cite{Electronic_imaging, zerman2020textured, VVDB2}, in which subjective experiments are conducted on eight different point cloud contents, each compressed by different encoder parameters, and a total of 152 compressed bit streams is generated. Using the subjective scores (i.e., mean opinion score, MOS) provided with the VSENSE VVDB2 database, we calculate a set of performance metrics for the objective scores measured by the PC-PSNR. These performance metrics include the Pearson correlation coefficient (PCC), Spearman rank-ordered correlation coefficient (SROCC), root mean square error (RMS), and the outlier ratio (OR) \cite{P.1401}. 
To evaluate the objective estimate, a non-linear logistic function \cite{Logistic_function} --i.e. $MOS_{pred} = \beta_2 + (\beta_1 - \beta_2)/({1 + e^{-(Q_{obj} - \beta_3)/|\beta_4|)}})$-- is employed for fitting the objective estimate to the subjective scores, and then the correlation coefficients are calculated between the predicted MOS and the collected MOS. The results are shown in Table \ref{Correlation-Coefficient}, where, for comparison, we also include the correlation coefficients of a variety of existing objective quality metrics, including point-to-point ($p2point$) \cite{Dong_Tian}, point-to-plane ($p2plane$) \cite{Dong_Tian}, and plane-to-plane ($pl2plane$) \cite{plane-to-plane} in the forms of $RMS$ (or $PSNR_{RMS}$), $Haus$ (or $PSNR_{Haus}$), or mean. The metrics also include the color difference between the closest corresponding points for $Y$, $U$, $V$ color channels and combined ($YUV$) in the forms of $MSE$ or $PSNR$. 
PCQM \cite{PCQM} is a recently developed overall quality metric to assess the visual quality of a colored 3D point cloud, which uses average weights to linearly combine the geometry-based and color-based features. 
{As can be observed from Table \ref{Correlation-Coefficient}, compared to the best-performing geometry distortion metric $p2point_{RMS}$, our unified model improves the PCC and SROCC coefficients by 0.013 and 0.088, respectively.  Compared to the best performing color distortion metric $PSNR_Y$, we improve the PCC and SROCC coefficients by 0.016 and 0.015, respectively. For RMS and OR indicators, the values of them are decreased to 7.59 and 0.031, respectively, by our proposed PC-PSNR.} 

}

\begin{figure*}[!htbp]
	\hspace{-1mm}\subfigure[AxeGuy]{
		\label{Fig.sub.1}
		\includegraphics[width=0.45\textwidth]{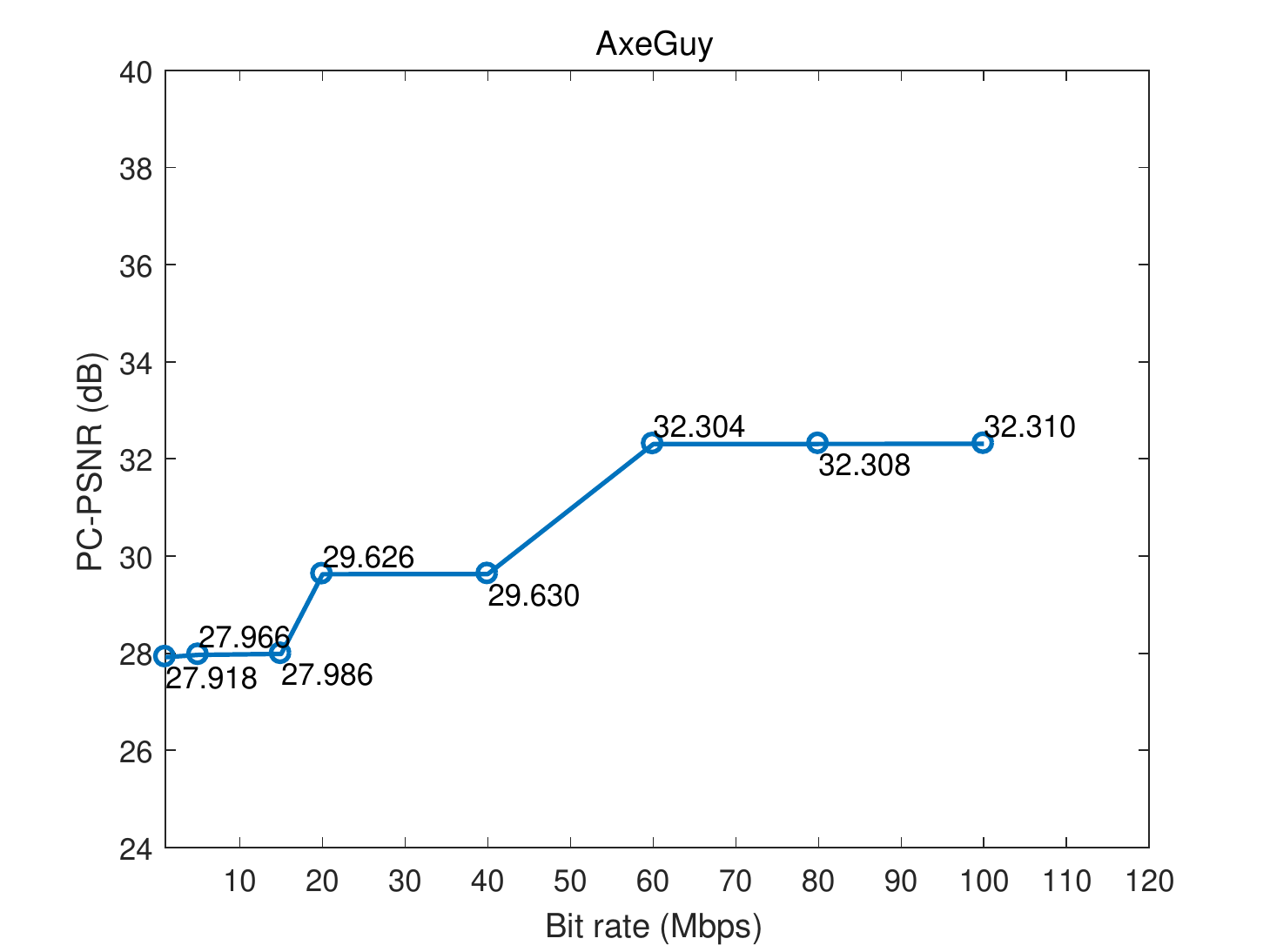}}
	\subfigure[Longdress]{
		\label{Fig.sub.1}
		\includegraphics[width=0.45\textwidth]{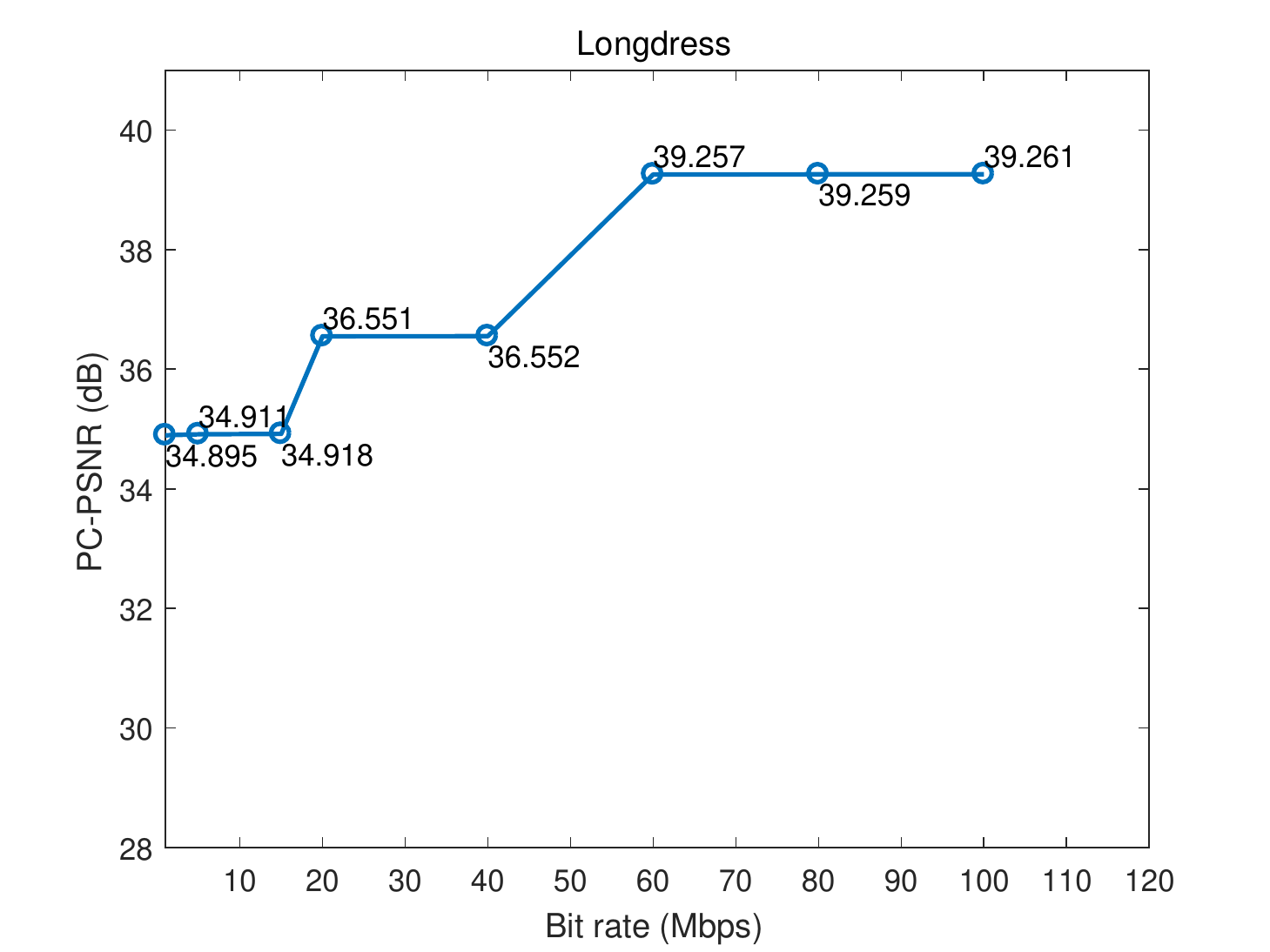}}
	
	\subfigure[Loot]{
		\label{Fig.sub.1}
		\includegraphics[width=0.45\textwidth]{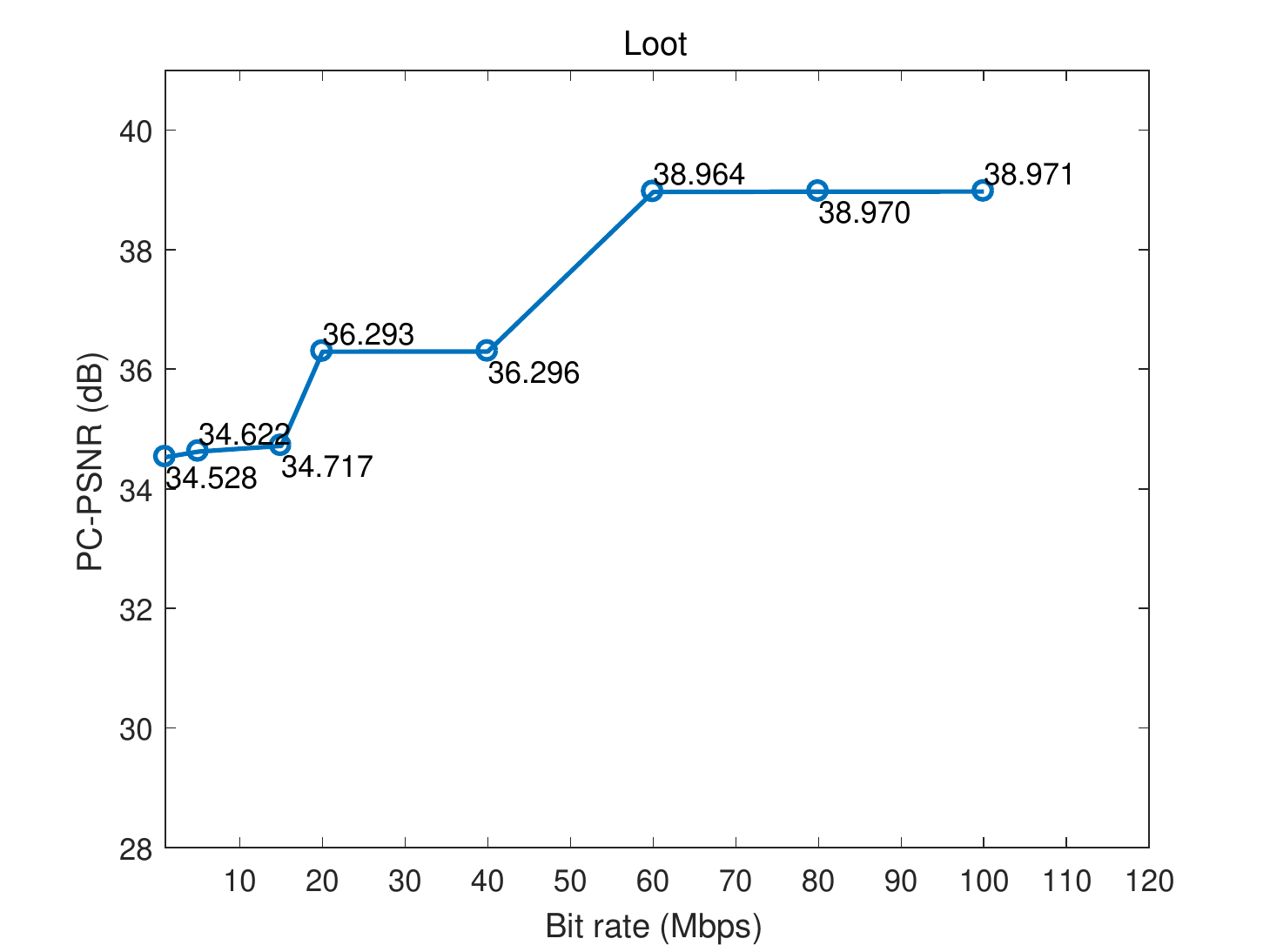}}
	\subfigure[Soldier]{
		\label{Fig.sub.1}
		\includegraphics[width=0.45\textwidth]{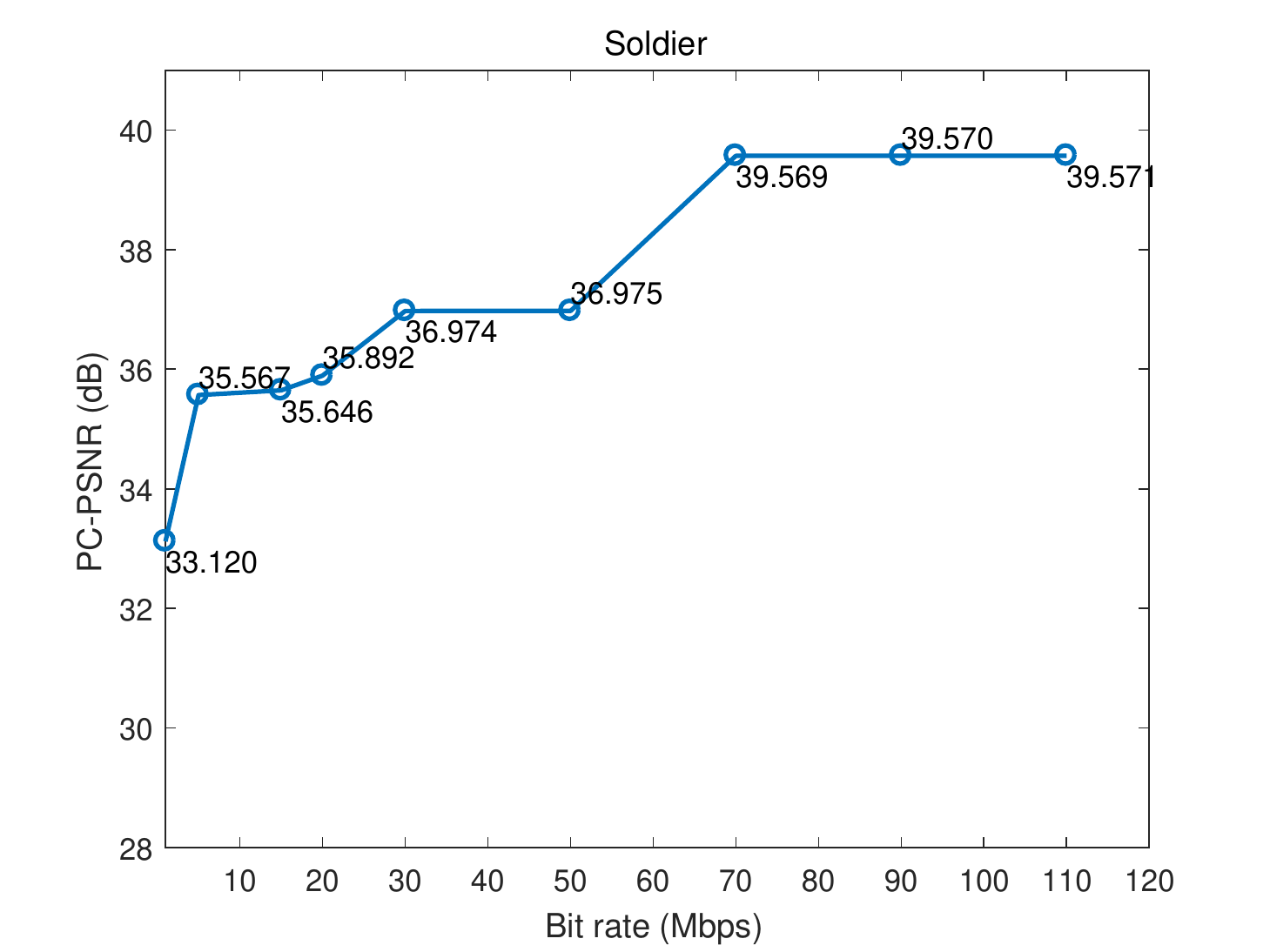}}
	
	\subfigure[Richardo10]{
		\label{Fig.sub.1}
		\includegraphics[width=0.45\textwidth]{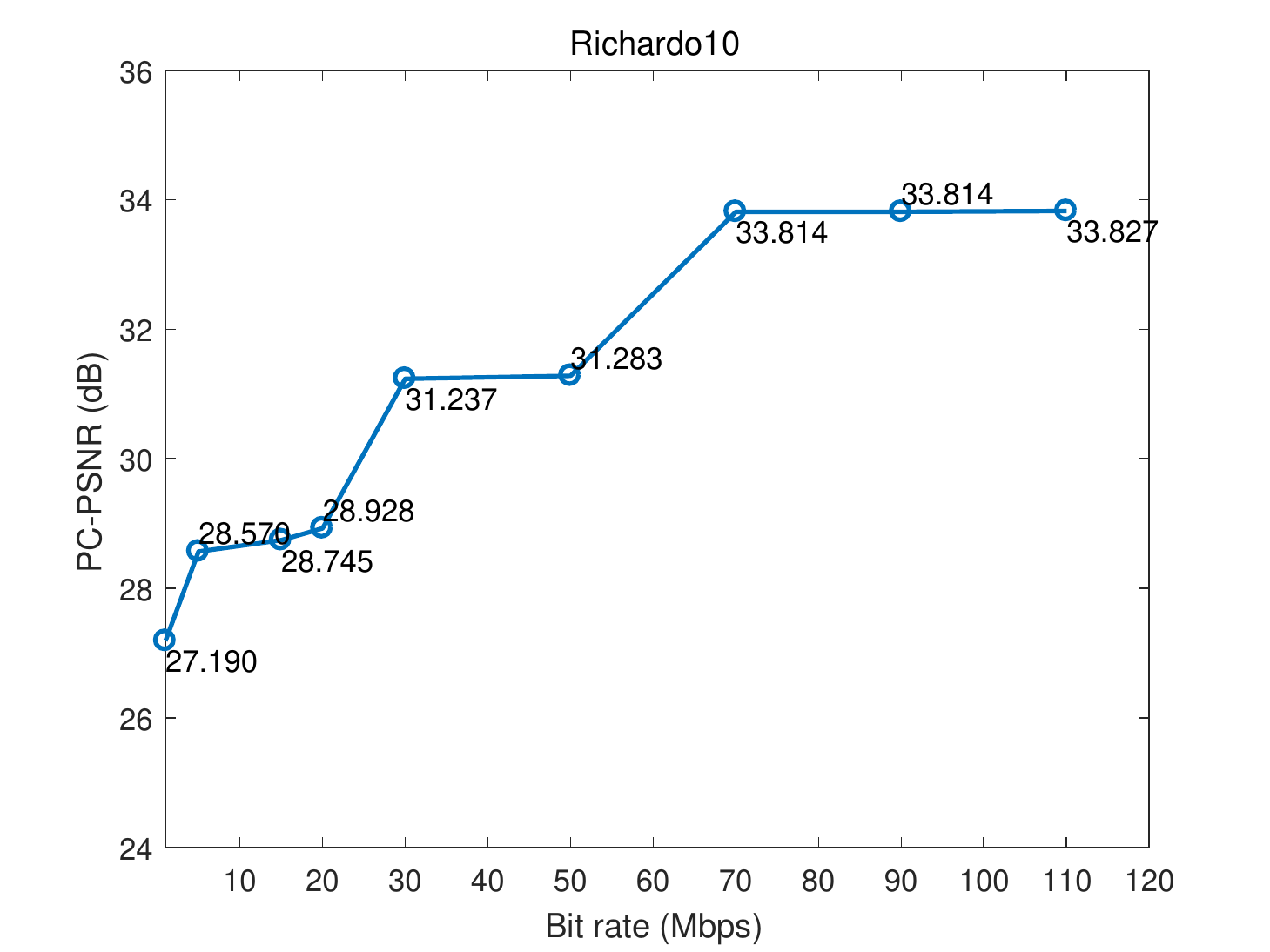}}
	\subfigure[Phil9]{
		\label{Fig.sub.1}
		\includegraphics[width=0.45\textwidth]{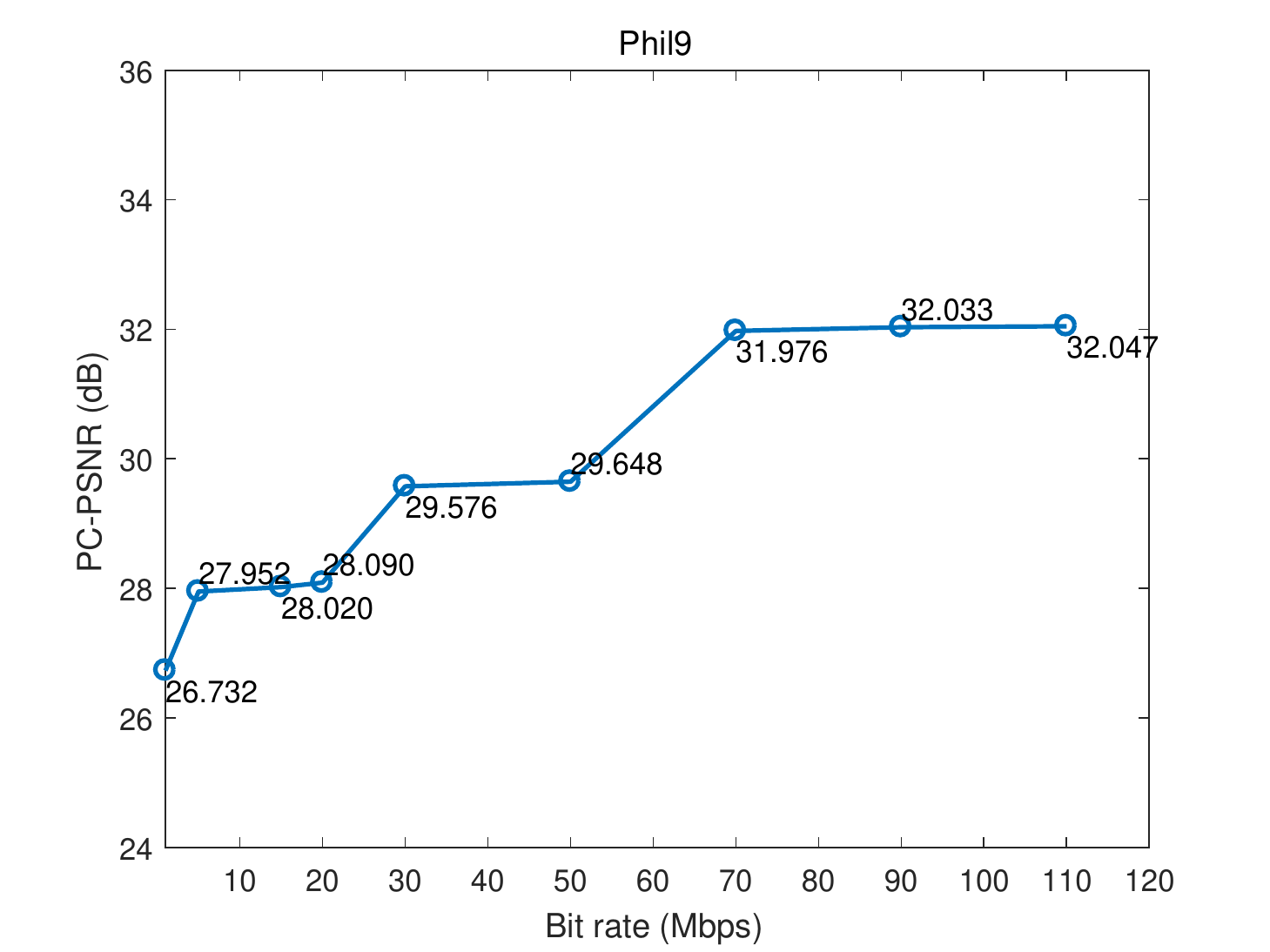}}
	
	\caption{{PC-PSNR versus target bit rate for the point clouds, where for better illustration, the specific PC-PSNR value is shown besides the collected rate-distortion data point in the graph. (a) AxeGuy. (b) Longdress. (c) Loot. (d) Soldier. (e) Richardo10. (f) Phil9.}}
	\label{rate-distortion-curve}
\end{figure*}

\begin{table*}[htbp]
	\scriptsize
	\centering
	\caption{Summary of the unified distortion   $D(\hat{\mathcal{P}},{\mathcal{P}})$, geometry distortion $d_g(\hat {\mathcal{P}}, {\mathcal{P}})$, and color distortion $d_{c}(\hat {\mathcal{P}}, {\mathcal{P}})$ of the compressed point clouds under the target bit rate.}
	\begin{tabular}{|l|l|l|l|l|l|l|l|l|l|}
		\hline
		\multicolumn{1}{|c|}{\multirow{2}{*}{\shortstack[l]{Bit rate \\ (Mbps)}}} & \multicolumn{3}{c|}{AxeGuy}   & \multicolumn{3}{c|}{Longdress}  & \multicolumn{3}{c|}{Loot}    \\ \cline{2-10} 
		\multicolumn{1}{|c|}{} & $D(\hat{\mathcal{P}},{\mathcal{P}})$ & $d_g(\hat {\mathcal{P}}, {\mathcal{P}})$ & $d_{c}(\hat {\mathcal{P}}, {\mathcal{P}})$ & $D(\hat{\mathcal{P}},{\mathcal{P}})$ & $d_g(\hat {\mathcal{P}}, {\mathcal{P}})$ & $d_{c}(\hat {\mathcal{P}}, {\mathcal{P}})$ & $D(\hat{\mathcal{P}},{\mathcal{P}})$ & $d_g(\hat {\mathcal{P}}, {\mathcal{P}})$ & $d_{c}(\hat {\mathcal{P}}, {\mathcal{P}})$ \\ \hline
		20                     & $4.360\times10^{-3}$ & $2.26\times10^{-1}$ & $8.93\times10^{-4}$    & $8.850\times10^{-4}$ & $7.34\times10^{-2}$ & $1.38\times10^{-3}$    & $9.392\times10^{-4}$ & $7.75\times 10^{-2}$ & $4.36\times10^{-4}$     \\ \hline
		40                     & $4.356\times10^{-3}$ & $2.24\times10^{-1}$ & $5.61\times10^{-4}$    & $8.848\times10^{-4}$ & $7.31\times10^{-2}$ & $7.63\times10^{-4}$    & $9.386\times10^{-4}$ & $7.74\times 10^{-2}$ & $2.52\times10^{-4}$     \\ \hline
		60                     & $2.443\times10^{-3}$ & $1.35\times10^{-1}$ & $3.39\times10^{-4}$    & $4.746\times10^{-4}$ & $4.21\times10^{-2}$ & $3.40\times10^{-4}$    & $5.078\times10^{-4}$ & $4.94\times 10^{-2}$ & $1.23\times10^{-4}$     \\ \hline
		80                     & $2.441\times10^{-3}$ & $1.32\times10^{-1}$ & $2.13\times10^{-4}$    & $4.744\times10^{-4}$ & $4.18\times10^{-2}$ & $1.36\times10^{-4}$    & $5.071\times10^{-4}$ & $4.92\times 10^{-2}$ & $4.03\times10^{-5}$     \\ \hline
		100                    & $2.440\times10^{-3}$ & $1.29\times10^{-1}$ & $1.56\times10^{-4}$    & $4.742\times10^{-4}$ & $4.15\times10^{-2}$ & $5.59\times10^{-5}$    & $5.069\times10^{-4}$ & $4.90\times 10^{-2}$ & $2.13\times10^{-5}$     \\ \hline
		\multicolumn{1}{|c|}{\multirow{2}{*}{\shortstack[l]{Bit rate \\ (Mbps)}}} & \multicolumn{3}{c|}{Soldier}   & \multicolumn{3}{c|}{Ricardo10}  & \multicolumn{3}{c|}{Phil9}    \\ \cline{2-10} 
		\multicolumn{1}{|c|}{} & $D(\hat{\mathcal{P}},{\mathcal{P}})$ & $d_g(\hat {\mathcal{P}}, {\mathcal{P}})$ & $d_{c}(\hat {\mathcal{P}}, {\mathcal{P}})$ & $D(\hat{\mathcal{P}},{\mathcal{P}})$ & $d_g(\hat {\mathcal{P}}, {\mathcal{P}})$ & $d_{c}(\hat {\mathcal{P}}, {\mathcal{P}})$ & $D(\hat{\mathcal{P}},{\mathcal{P}})$ & $d_g(\hat {\mathcal{P}}, {\mathcal{P}})$ & $d_{c}(\hat {\mathcal{P}}, {\mathcal{P}})$ \\ \hline
		30                     & $8.029\times10^{-4}$ & $9.15\times10^{-2}$ & $7.87\times10^{-4}$    & $3.01\times10^{-3}$ & $2.472\times10^{-1}$ & $1.03\times10^{-4}$    & $4.41\times10^{-3}$ & $2.264\times 10^{-1}$ & $7.87\times10^{-4}$     \\ \hline
		50                     & $8.028\times10^{-4}$ & $8.99\times10^{-2}$ & $4.41\times10^{-4}$    & $2.98\times10^{-3}$ & $2.446\times10^{-1}$ & $7.32\times10^{-5}$    & $4.33\times10^{-3}$ & $2.227\times 10^{-1}$ & $5.48\times10^{-4}$     \\ \hline
		70                     & $4.417\times10^{-4}$ & $4.69\times10^{-2}$ & $2.09\times10^{-4}$    & $1.69\times10^{-3}$ & $1.735\times10^{-1}$ & $4.51\times10^{-5}$    & $2.53\times10^{-3}$ & $1.427\times 10^{-1}$ & $3.09\times10^{-4}$     \\ \hline
		90                    & $4.416\times10^{-4}$ & $4.67\times10^{-2}$ & $9.56\times10^{-5}$    & $1.67\times10^{-3}$ & $1.732\times10^{-1}$ & $2.99\times10^{-5}$    & $2.51\times10^{-3}$ & $1.410\times 10^{-1}$ & $2.44\times10^{-4}$     \\ \hline
		110                    & $4.415\times10^{-4}$ & $4.65\times10^{-2}$ & $4.30\times10^{-5}$    & $1.65\times10^{-3}$ & $1.730\times10^{-1}$ & $2.23\times10^{-5}$    & $2.49\times10^{-3}$ & $1.406\times 10^{-1}$ & $2.08\times10^{-4}$     \\ \hline
	\end{tabular}\label{Distortion_list_table}
\end{table*}

\subsection{Performance of the Proposed Rate Distortion Optimization Scheme}

\begin{figure*}[!htbp]
		\centering
	\hspace{-1mm}\subfigure[20 Mbps]{
		\label{Fig.sub.1}
		\includegraphics[width=0.23\textwidth]{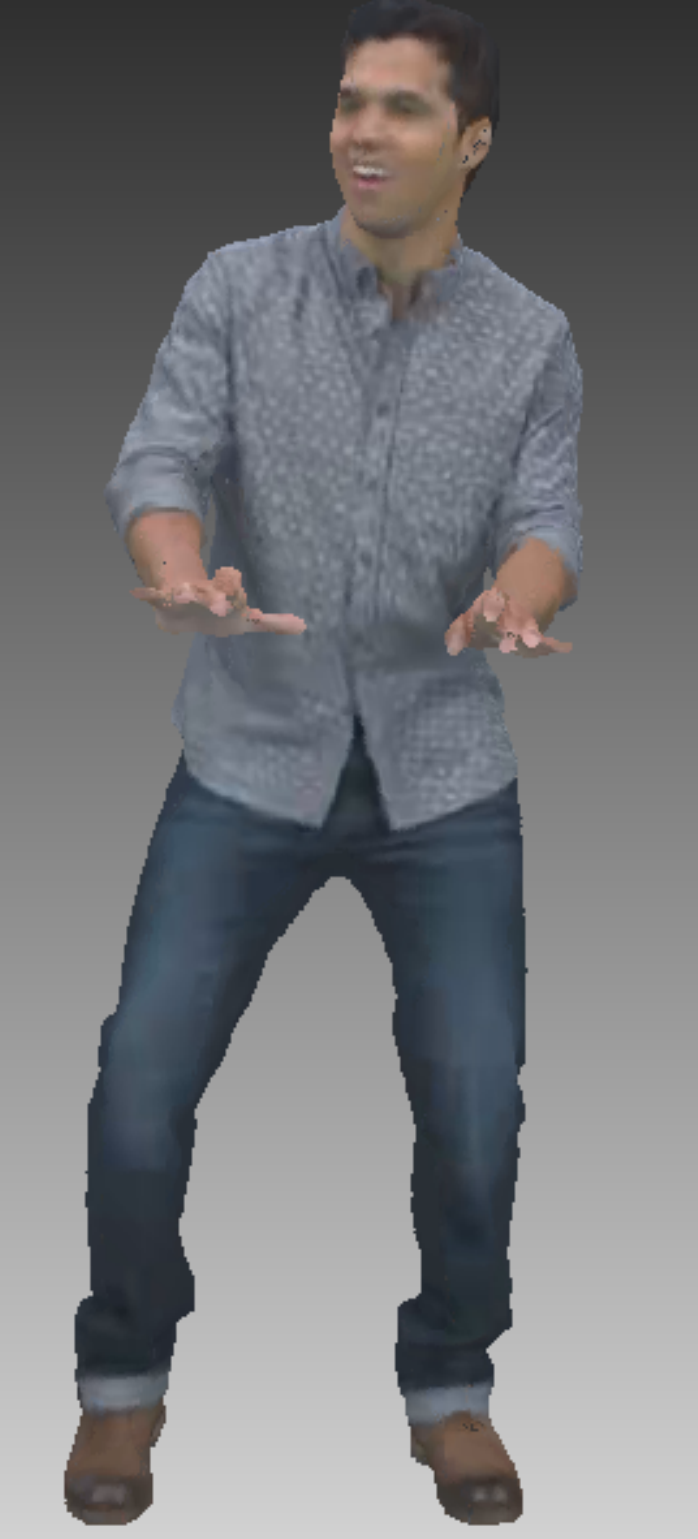}}
	\subfigure[40 Mbps]{
		\label{Fig.sub.1}
		\includegraphics[width=0.23\textwidth]{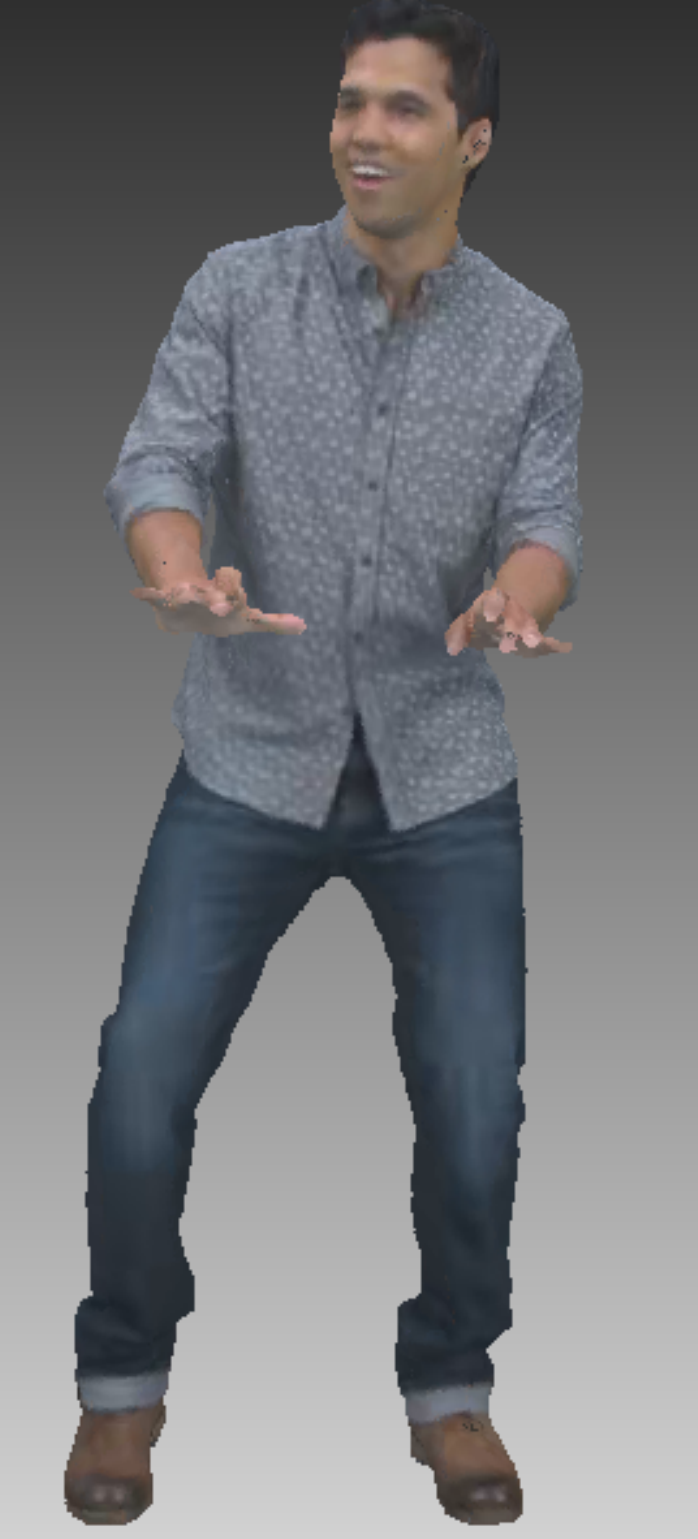}}
	\subfigure[60 Mbps]{
		\label{Fig.sub.1}
		\includegraphics[width=0.23\textwidth]{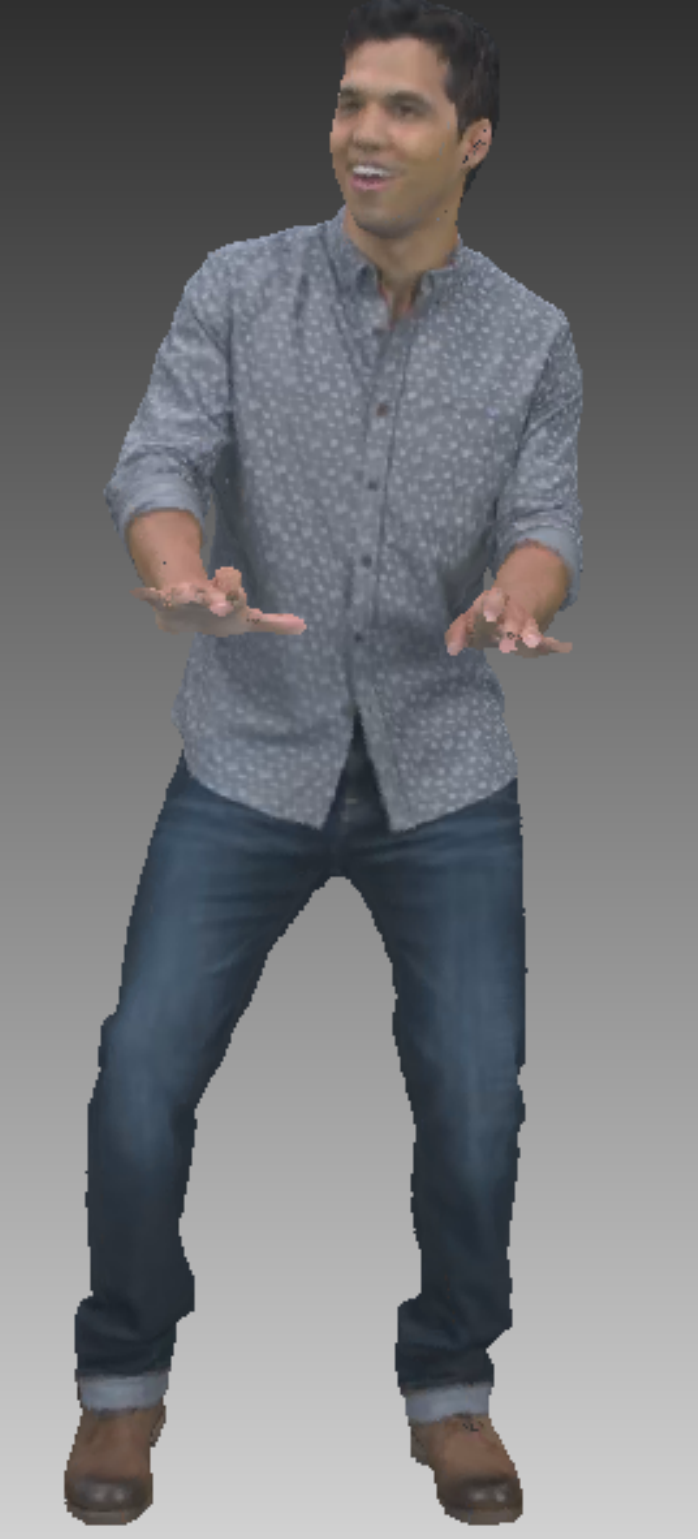}}
	\subfigure[80 Mbps]{
		\label{Fig.sub.1}
		\includegraphics[width=0.23\textwidth]{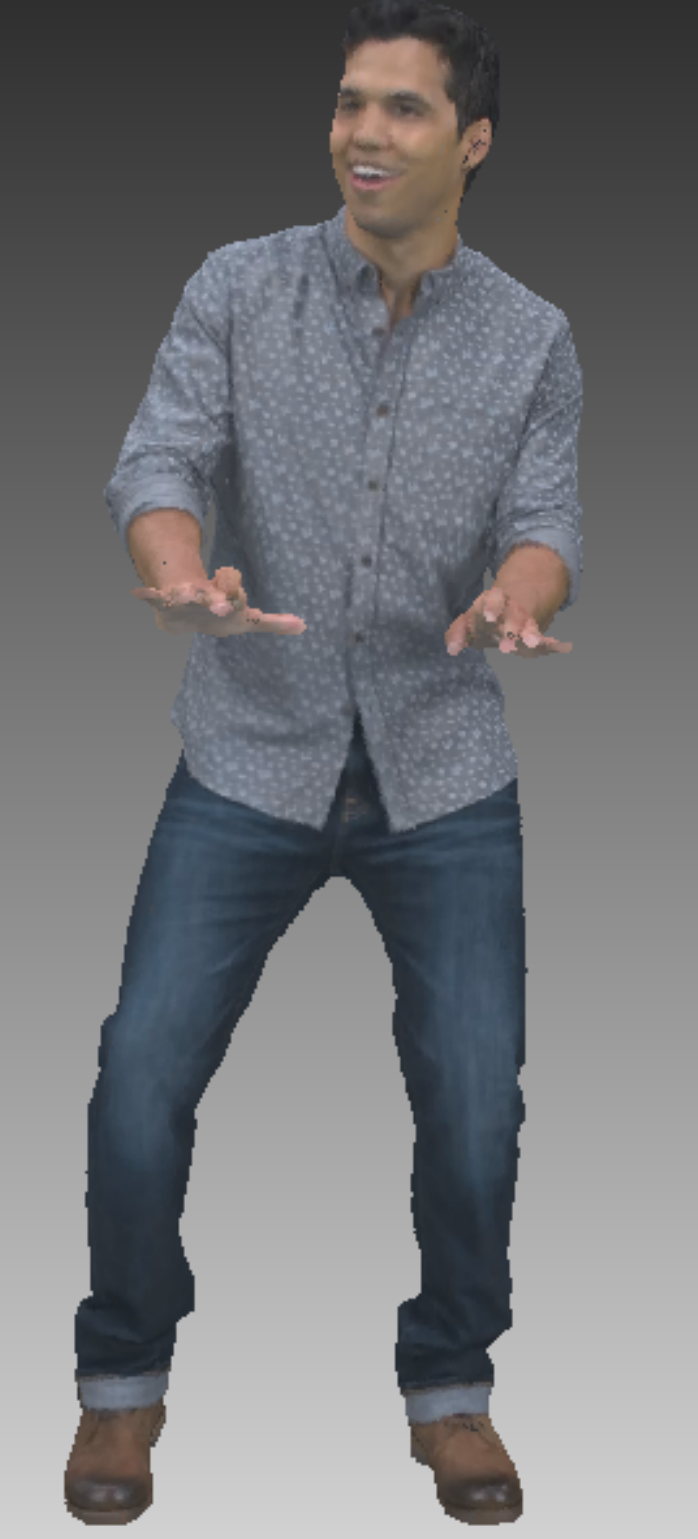}}
	
	\caption{Visual results of the point clouds coded with the proposed algorithm under various target bit rates. (a) Target bit rate: 20 Mbps. (b) Target bit rate: 40 Mbps. (c) Target bit rate: 60 Mbps. (d) Target bit rate: 80 Mbps.}
	\label{visual_point_clouds}
\end{figure*}

Fig. \ref{rate-distortion-curve} shows the rate-distortion performance of the proposed algorithm on various test point clouds. As can be observed, different from the rate-distortion curve for bit rate constrained coding of color image or video, which exhibits convex characteristics, the graph formed by connecting the rate-distortion points of point clouds is approximately piece-wise linear.  That is, for a certain range of bit rates, the rate distortion points constructs nearly a horizontal line.  In other words, the proposed algorithm achieves almost identical decoded quality for a number of target bit rates. To understand that, we calculate the  unified distortion, geometry distortion, and color distortion for the point clouds compressed by using the proposed algorithm under each target bit rate, the results of which are listed in the Table \ref{Distortion_list_table}. As can be observed, as the target bit rate changes from 100 Mbps to 60 Mbps, the resulting geometry distortion is kept unchanged, and the color distortion increases linearly. However, the unified distortion almost remains the same (although the unified distortion also increases with the decrease of the bit rate, the increment at each bit rate change step is very little.). This means that, the color distortion  contributes little to the proposed unified distortion that is calculated by combining the geometry and color distortions using the covariance matrix of the point clouds. In order to verify this conjecture, we conduct further experiments to derive the gradients of unified distortion with respect to geometry QP and color QP, which are represented as $\nabla_{q_g}D(\hat{\mathcal{P}},{\mathcal{P}})$ and $\nabla_{q_c}D(\hat{\mathcal{P}},{\mathcal{P}})$, respectively. In the calculation of $\nabla_{q_g}D(\hat{\mathcal{P}},{\mathcal{P}})$, we fix bit rate  for color coding (i.e., fixing QP for color), and vary the bit rate for geometry coding (i.e., varying QP for geometry), and then calculate the corresponding unified distortion change to calculate the gradient. We use similar method to numerically calculate $\nabla_{q_c}D(\hat{\mathcal{P}},{\mathcal{P}})$. The results are given in the Table \ref{gradient_qp}. From this table, it is easy to see that, $\nabla_{q_g}D(\hat{\mathcal{P}},{\mathcal{P}})$ is much larger than $\nabla_{q_c}D(\hat{\mathcal{P}},{\mathcal{P}})$, which means that the unified distortion is indeed insensitive to the change of color QP. This explains why   $D(\hat{\mathcal{P}},{\mathcal{P}})$ changes a little when  $d_{c}(\hat {\mathcal{P}}, {\mathcal{P}})$ changes in Table \ref{Distortion_list_table}.

On the other hand, Tables \ref{Distortion_list_table} and \ref{gradient_qp} also demonstrate the effectiveness of the proposed bit-rate constrained point cloud compression algorithm. Since the geometry distortion contributes significantly more than color distortion to the unified distortion, we should increase the geometry reconstruction quality as much as possible when encoding the point clouds. As shown in Table \ref{Distortion_list_table}, the geometry distortions are generally small for all the chosen target bit rates. 
It is worth pointing out that, due to different scales used for geometry coordinate and color attribute, geometry distortion is still much larger than color distortion. 
In order better to see how the proposed algorithm works, we provide the final geometry and color QPs selected by our proposed method and the corresponding bit rate allocation results in Table \ref{QP_display}. 
From this table, we  found that, the geometry QPs used for coding after employing our proposed algorithm  is usually within the range of $[2,10]$ for the test point clouds. This reveals that the proposed rate-distortion modeling algorithm can indeed choose an appropriate geometry QP to protect the geometry quality. As the target bit rate decreases from 110 Mbps to 70 Mbps, the color distortion increases, and geometry distortion (geometry QP) is not changed too much. This means that, in response to the target bit rate decrease, the proposed algorithm mainly relies on increasing the QP of color coding. Since color distortion will not affect the unified distortion too much,  this rate-distortion optimization  guarantees that good overall quality of point clouds can be achieved for all the target bit rates. This thus explains why the rate-distortion curve of the proposed algorithm for point cloud compression is an approximate straight line for a certain range of bit rates.

Fig. \ref{visual_point_clouds} shows the decoded subjective quality of the  point clouds  under the bit rates of 20, 40, 60, 80 Mbps coded with the proposed algorithm. Generally speaking, the proposed algorithm  achieves very good overall visual quality for all target bit rates. However, in the low bit rate case, due to very few bit rate allocated for color coding (e.g., only 1.6 Mbps is allocated for color information in coding of soldier when target bit rate is 30 Mbps as shown in Table \ref{QP_display}), some blur effect can be inevitably observed in the resulting point clouds, i.e., Fig. \ref{visual_point_clouds}(a). Nevertheless,  compared to geometry distortion inducing holes, this artifact is perceptually acceptable.

\subsection{Comparison with the Existing Model-Based Optimization Scheme}
To further demonstrate the effectiveness of the proposed algorithm, we would like to make a comparison of the proposed algorithm with other algorithms. However, to the best of our knowledges, there is very limited work in the literature that addresses the problem of bit rate allocation for video-based point cloud compression. Instead, to do a comparison, we firstly choose to extend another existing bit rate allocation scheme, which was originally proposed for 3D mesh compression, to point cloud coding. For this purpose, we choose the model-based 3D mesh rate allocation scheme proposed in \cite{rate-distortion-model-3D-mesh} as the basis for extension to point cloud coding. In \cite{rate-distortion-model-3D-mesh}, 3D dynamic mesh (i.e., 3D facial expressions) is converted into geometry video, and two exponential functions are fitted to model the relationships of the distortion and rate of the geometry video with the QP. By using the Lagrange multiplier method, a closed-form solution of geometry QP is derived for 3D mesh coding. As can be observed, this scheme is using rate and distortion models characterized on the geometry video for formulation of the bit rate allocation problem for 3D mesh.

Following this idea, we design the reference algorithm for point cloud compression as follows. We firstly use the V-PCC, only to obtain the texture video and geometry video from the point clouds. This is done to simulate the first step proposed for the 3D mesh compression scheme~\cite{rate-distortion-model-3D-mesh}. Then, we model the distortion-quantization and rate-quantization relationships for the texture and geometry videos using the exponential functions. We sum the distortions of the texture video and geometry video as final distortion, and the rates of the texture video and geometry video as final bit rate for point clouds. As the distortion and rate models are established on the planar videos, we can derive the optimal QPs for geometry coding and texture coding using the numerical method. This Video-Rate-Distortion Modeled Point Cloud Compression algorithm is referred to as ``VRDM\_PCC'' for brevity.

\begin{table}[tbp]
	\centering
	\caption{Gradients of the unified distortion with geometry and color QPs.}
	\begin{tabular}{l|l|l}
		\hline
		Sequence  & $\nabla_{q_g}D(\hat{\mathcal{P}},{\mathcal{P}})$ & $\nabla_{q_c}D(\hat{\mathcal{P}},{\mathcal{P}})$ \\ \hline
		AxeGuy  & $2.5\times10^{-4}$                                             & $8\times10^{-7}$                                           \\ \hline
		Longdress & $2.0\times10^{-4}$                                            & $0.2\times10^{-7}$                                            \\ \hline
		Loot      & $1.8\times10^{-4}$                                            & $1.2\times10^{-7}$                                           \\ \hline
		Soldier   & $2.2\times10^{-4}$                                           & $0.08\times10^{-7}$                                          \\ \hline
		Richardo10     & $1.3\times10^{-3}$                                            & $4.0\times10^{-7}$                                           \\ \hline
		Phil9   & $3.5\times10^{-3}$                                           & $1.0\times10^{-7}$                                          \\ \hline
	\end{tabular}\label{gradient_qp}
\end{table}

{\renewcommand\baselinestretch{1.2}\selectfont
	\begin{table}[htbp]
		\centering
		\renewcommand{\arraystretch}{1.1}
		\caption{{QPs selected  by the proposed algorithm and associated bit rate allocation results (Mbps), where $q_c$ (or $q_g$) is the quantization parameter, and  $R(q_c)$ (or $R(q_g)$) is the bit rate.}}
		\label{QP_display}
		\begin{tabular}{|p{0.8cm}<{\centering}|p{0.4cm}<{\centering}|p{0.4cm}<{\centering}|p{0.6cm}<{\centering}|p{0.6cm}<{\centering}|p{0.4cm}<{\centering}|p{0.4cm}<{\centering}|p{0.6cm}<{\centering}|p{0.6cm}<{\centering}|}
			\hline
			\multicolumn{1}{|c|}{\multirow{3}{*}{\begin{tabular}[c]{@{}c@{}}Bit rate\\ (Mbps)\end{tabular}}} & \multicolumn{8}{c|}{QP and bit rate allocation results}                              \\ \cline{2-9} 
			\multicolumn{1}{|c|}{}                                                                           & \multicolumn{4}{c|}{Soldier} & \multicolumn{4}{c|}{Richardo10}\\ \cline{2-9} 
			\multicolumn{1}{|c|}{}                                                                           & $q_c$         & $q_g$        & $R(q_c)$          & $R(q_g)$          & $q_c$        & $q_g$    & $R(q_c)$          & $R(q_g)$     \\ \hline
			30                                                                                               & 39            & 10          & 1.6            & 28.4   & 29            & 9              & 2.3           & 27.7          \\ \hline
			50                                                                                               & 29            & 9        & 4.2            & 45.8     &25             & 8              & 3.4           & 46.6          \\ \hline
			70                                                                                               & 23            & 4      & 9.7            & 60.3      &20             & 3              & 5          & 65           \\ \hline
			90                                                                                               & 15            & 3      & 24            & 66      & 13             & 3              & 11.8          & 78.2          \\ \hline
			110                                                                                              & 9           & 2      & 45.5            & 64.5      & 6             & 2              & 34.7           & 75.3           \\ \hline
		\end{tabular}
	\end{table}
}

 \begin{figure}[!htbp]
 	\centering
 	\hspace{-1mm}\subfigure[Longdress - VRDM\_PCC]{
 		\label{Fig.sub.1}
 		\includegraphics[width=0.23\textwidth]{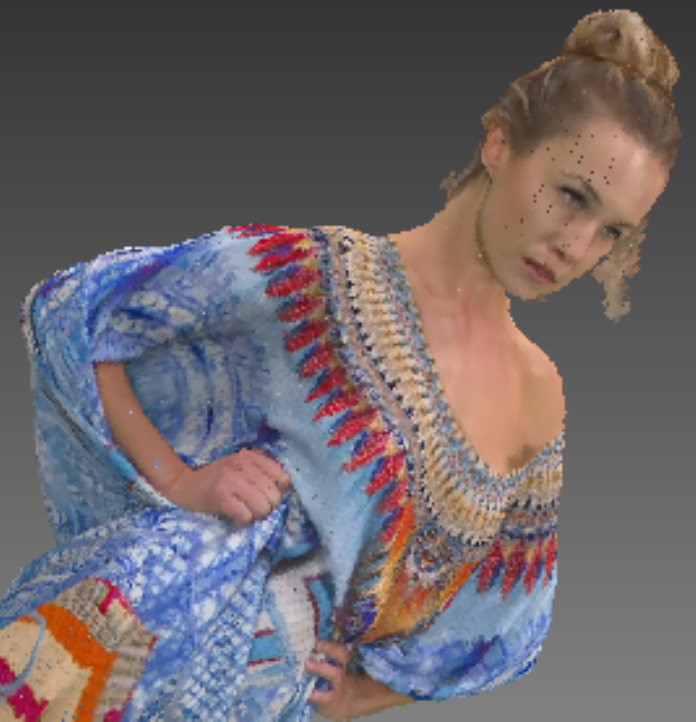}}
 	\subfigure[Longdress - Proposed]{
 		\label{Fig.sub.1}
 		\includegraphics[width=0.23\textwidth]{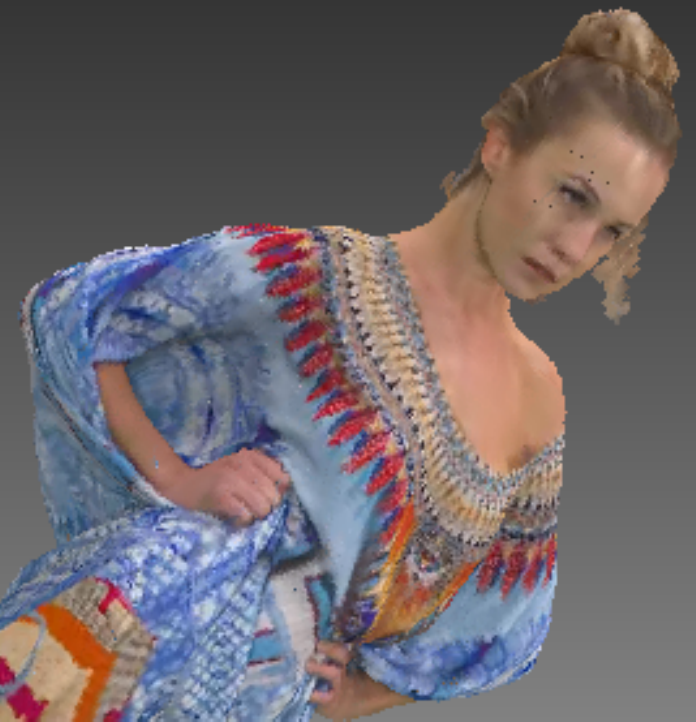}}
 	
 	\hspace{-1mm}\subfigure[Loot - VRDM\_PCC]{
 		\label{Fig.sub.1}
 		\includegraphics[width=0.23\textwidth]{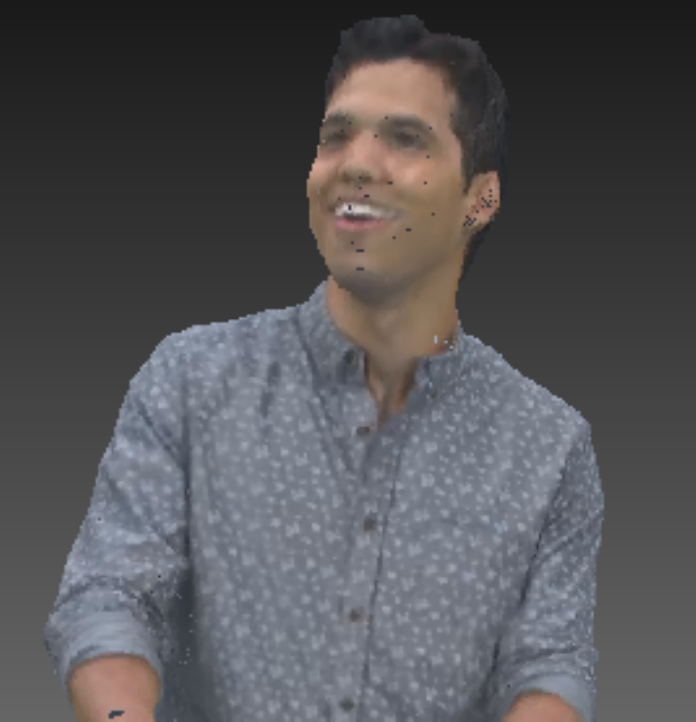}}
 	\subfigure[Loot - Proposed]{
 		\label{Fig.sub.1}
 		\includegraphics[width=0.23\textwidth]{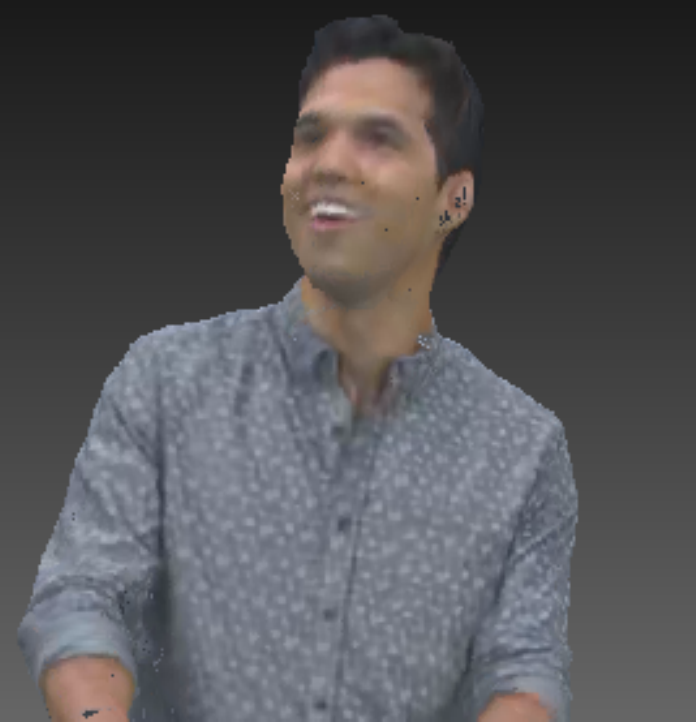}}
 	\caption{Subjective quality comparison between the proposed algorithm and the VRDM\_PCC algorithm. (a) Longdress frame coded with the VRDM\_PCC algorithm. (b)  Longdress frame coded with the proposed algorithm. (c) Loot frame coded with the VRDM\_PCC algorithm. (d) Loot frame coded with the proposed algorithm.}
 	\label{visual_VRDM}
 	\vspace{-5mm}
 \end{figure}

 \begin{table*}[!htbp]
	\scriptsize
	\caption{Objective performance comparison between the proposed algorithm and VRDM\_PCC under the target bit rate of 40 or 50 Mbps.}
	\hspace{-1mm}
	\centering
	\begin{tabular}{|l|l|l|l|l|l|l|l|l|}
		\hline
		\multicolumn{1}{|c|}{\multirow{2}{*}{Metric}}   & \multicolumn{2}{c|}{Longdress (40 Mbps)}             & \multicolumn{2}{c|}{Loot (40 Mbps)}          & \multicolumn{2}{c|}{Ricardo10 (50 Mbps)}             & \multicolumn{2}{c|}{Phil9 (50 Mbps)}             \\ \cline{2-9} 
		\multicolumn{1}{|c|}{}                          & VRDM\_PCC           & Proposed             & VRDM\_PCC           & Proposed    & VRDM\_PCC           & Proposed             & VRDM\_PCC           & Proposed         \\ \hline
		PC-PSNR                                         & 31.38               & 36.55                & 31.59               & 36.29    & 29.83               & 31.28               & 27.51              & 29.64            \\ \hline
		$D(\hat{\mathcal{P}},{\mathcal{P}})$            & $2.91\times10^{-3}$ & $8.84\times10^{-4}$  & $2.77\times10^{-3}$ & $9.38\times10^{-4}$ & $4.16\times10^{-3}$ & $2.98\times10^{-3}$  & $7.09\times10^{-3}$ & $4.33\times10^{-3}$ \\ \hline
		$d_g(\hat {\mathcal{P}}, {\mathcal{P}})$        & $24.0\times10^{-2}$ & $7.31\times10^{-2}$  & $22.9\times10^{-2}$ & $7.74\times 10^{-2}$ & $3.415\times10^{-1}$ & $2.446\times10^{-1}$  & $3.65\times10^{-1}$ & $2.227\times10^{-1}$ \\ \hline
		$d_{c}(\hat {\mathcal{P}}, {\mathcal{P}})$      & $3.24\times10^{-4}$ & $7.63\times10^{-4}$  & $9.35\times10^{-5}$ & $2.52\times10^{-4}$ & $4.06\times10^{-5}$ & $7.32\times10^{-5}$  & $3.54\times10^{-4}$ & $5.48\times10^{-4}$ \\ \hline
	\end{tabular}\label{objective_performance_VRDM_PCC}
\end{table*}

Fig. \ref{visual_VRDM} illustrates the subjective quality comparison for the point cloud frame between the proposed algorithm and the VRDM\_PCC. As can be observed from Fig. \ref{visual_VRDM}, compared to the proposed algorithm, the point cloud frame coded with VRDM\_PCC has much more black dots (i.e., holes) on the face of the human subject, which will lead to the rendered view more  unpleasant.  This demonstrates the proposed algorithm outperforms the VRDM\_PCC visually. The associated objective quality comparison between these two algorithms is tabulated in Table \ref{objective_performance_VRDM_PCC}. It can be seen that, in terms of PC-PSNR, the proposed algorithm yields an average gain of about 3.3 dB.  Therefore, we can conclude that the proposed algorithm produces better performance than VRDM\_PCC. The reason behind that can be explained as follows. 
In VRDM\_PCC, the rate and distortion models characterized on the generated videos are employed for determining the encoding QPs for point clouds, which neglects the effect of parametrization/mapping on the point cloud distortion. Instead, our proposed algorithm  models the color distortion and geometry distortion directly on the point clouds, and combines them using a unified model. With the unified model to balance the color distortion and geometry distortion, the proposed algorithm can find the optimal color and geometry QPs for bit rate constrained point cloud compression.

Relatively recently, to address the problem of bit allocation between geometry and color in video-based point cloud compression, a similar model-based rate allocation scheme is proposed in \cite{model-based-scheme}, which is considered as the first and also the state-of-the-art work in this area. In this scheme, the overall distortion of the point clouds  is firstly modeled as a linear combination of geometry distortion and color distortion using a weighting factor. The geometry distortion is measured using the point-to-point metric. Then, the function of the overall distortion with geometry and color QP is modeled as linear, and the relationship between bit rate and QPs is expressed as exponential. Finally, an interior point method is used to solve the constrained bit rate allocation problem. In the implementation, following the parameter setting in \cite{model-based-scheme}, we set the factor used for weighting geometry and color distortions to 0.5, and  use the established overall distortion and rate models to obtain geometry and color QPs for a given target bit rate. The performance  comparison between our proposed modeling algorithm with this method is shown in Table \ref{Comparison_model_proposed}, where the target bit rate for AxeGuy, Longdress, Loot is 20 Mbps, and the target bit rate for Soldier, Ricardo10, Phil9 is 30 Mbps. As we can observe from the Table, our proposed algorithm significantly outperforms the scheme proposed in \cite{model-based-scheme}. The reason for the substantial quality improvement brought by our  algorithm is that  we explicitly consider  statistical correlation between geometry and color distortions when modeling the overall distortion, and selectively protect the geometry quality during optimization. This kind of balance between geometry and color distortions, however, is not studied in \cite{model-based-scheme}. 

\begin{table}[h]
	\centering
	\caption{Performance comparison between  our proposed algorithm and the scheme in \cite{model-based-scheme} in terms of PC-PSNR.}
	\begin{tabular}{l|l|l}
		\hline
		Sequence  & Scheme \cite{model-based-scheme}  & Proposed \\ \hline
		AxeGuy  & 27.61                                             & 29.62                                           \\ \hline
		Longdress & 33.14                                            &    36.55                                         \\ \hline
		Loot      & 32.78                                           &    36.29                                      \\ \hline
		Soldier   & 33.41                                           & 36.97                                         \\ \hline
		Richardo10     & 28.07                                           & 31.24                                          \\ \hline
		Phil9   & 27.41                                          & 29.58                                          \\ \hline
	\end{tabular}\label{Comparison_model_proposed}
\end{table}

\begin{figure}[!htbp]
		\centering
		\hspace{-1mm}\subfigure[Longdress]{
			\label{Fig.sub.1}
			\includegraphics[width=0.235\textwidth]{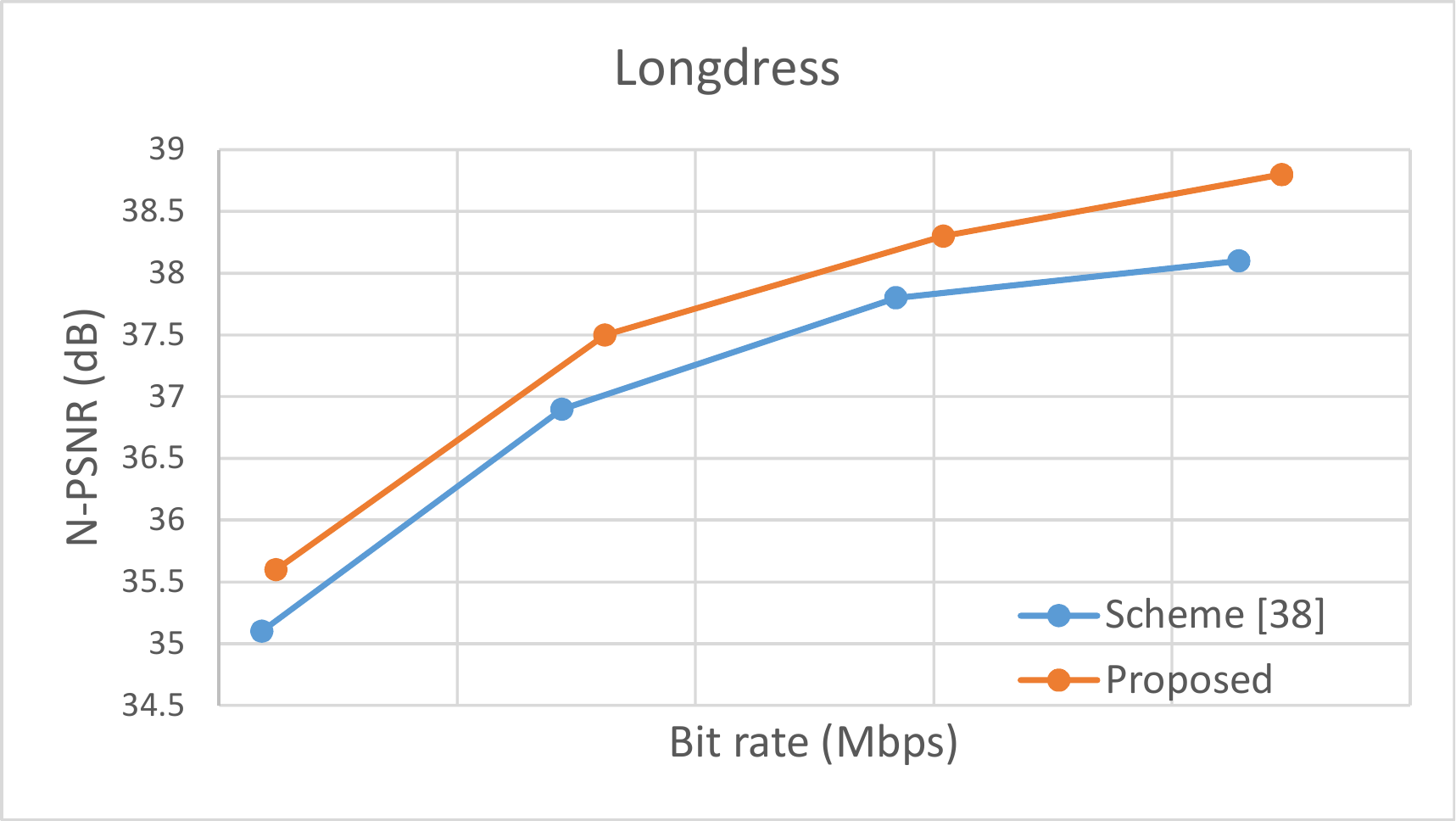}}
		\subfigure[Soldier]{
			\label{Fig.sub.1}
			\includegraphics[width=0.235\textwidth]{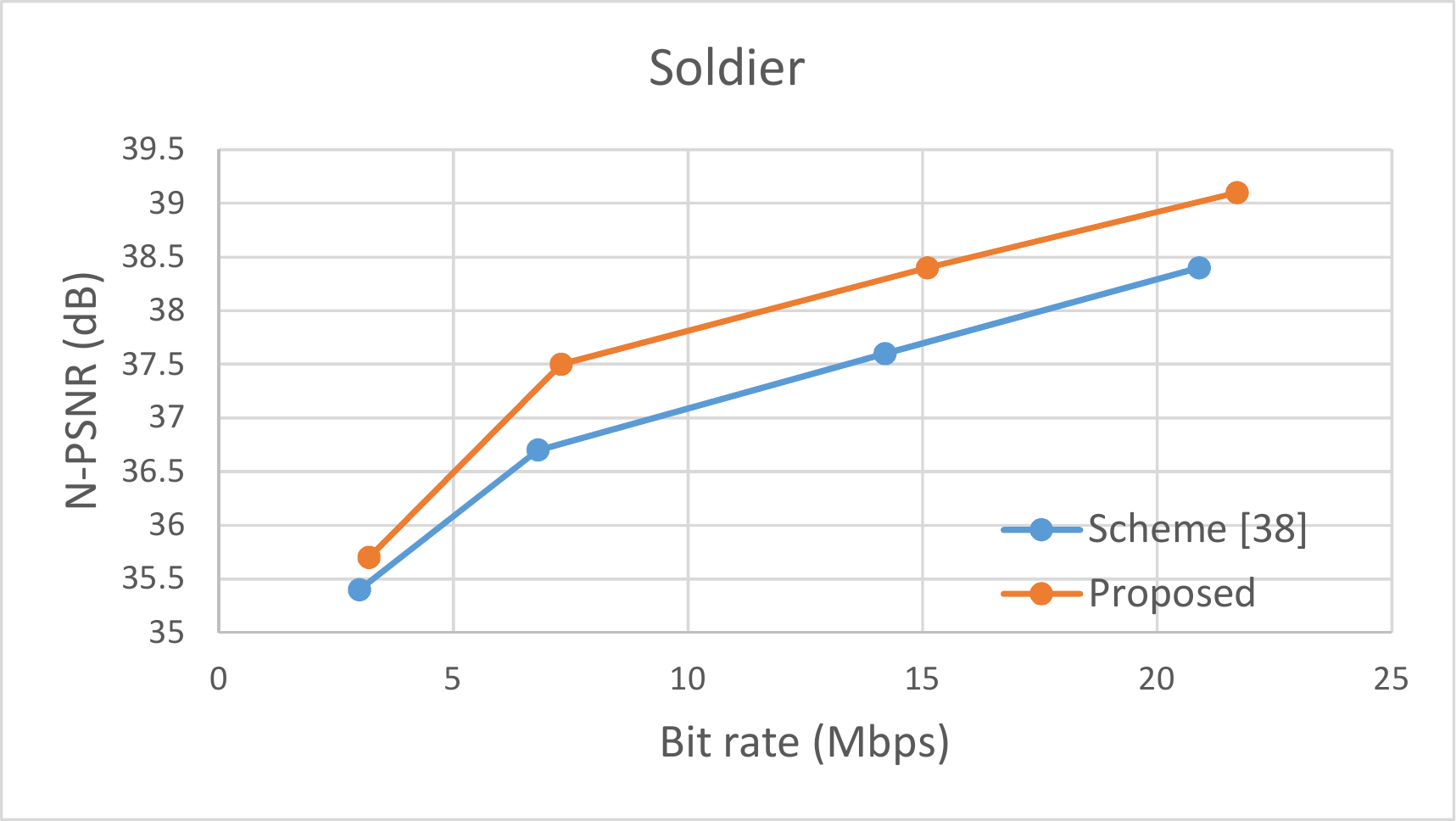}}
		
		\hspace{-1mm}\subfigure[Loot]{
			\label{Fig.sub.1}
			\includegraphics[width=0.235\textwidth]{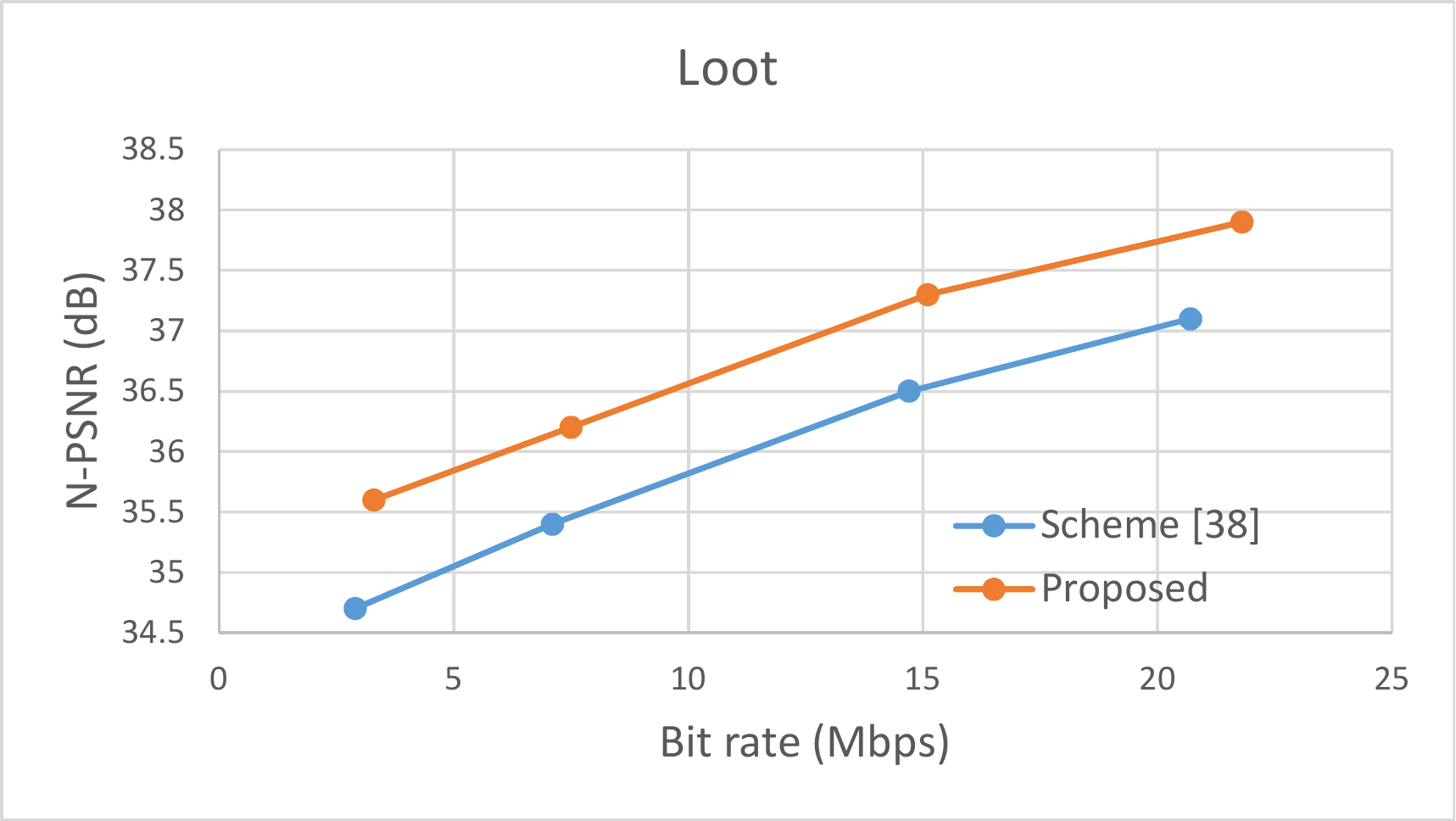}}
		\subfigure[Phil9]{
			\label{Fig.sub.1}
			\includegraphics[width=0.235\textwidth]{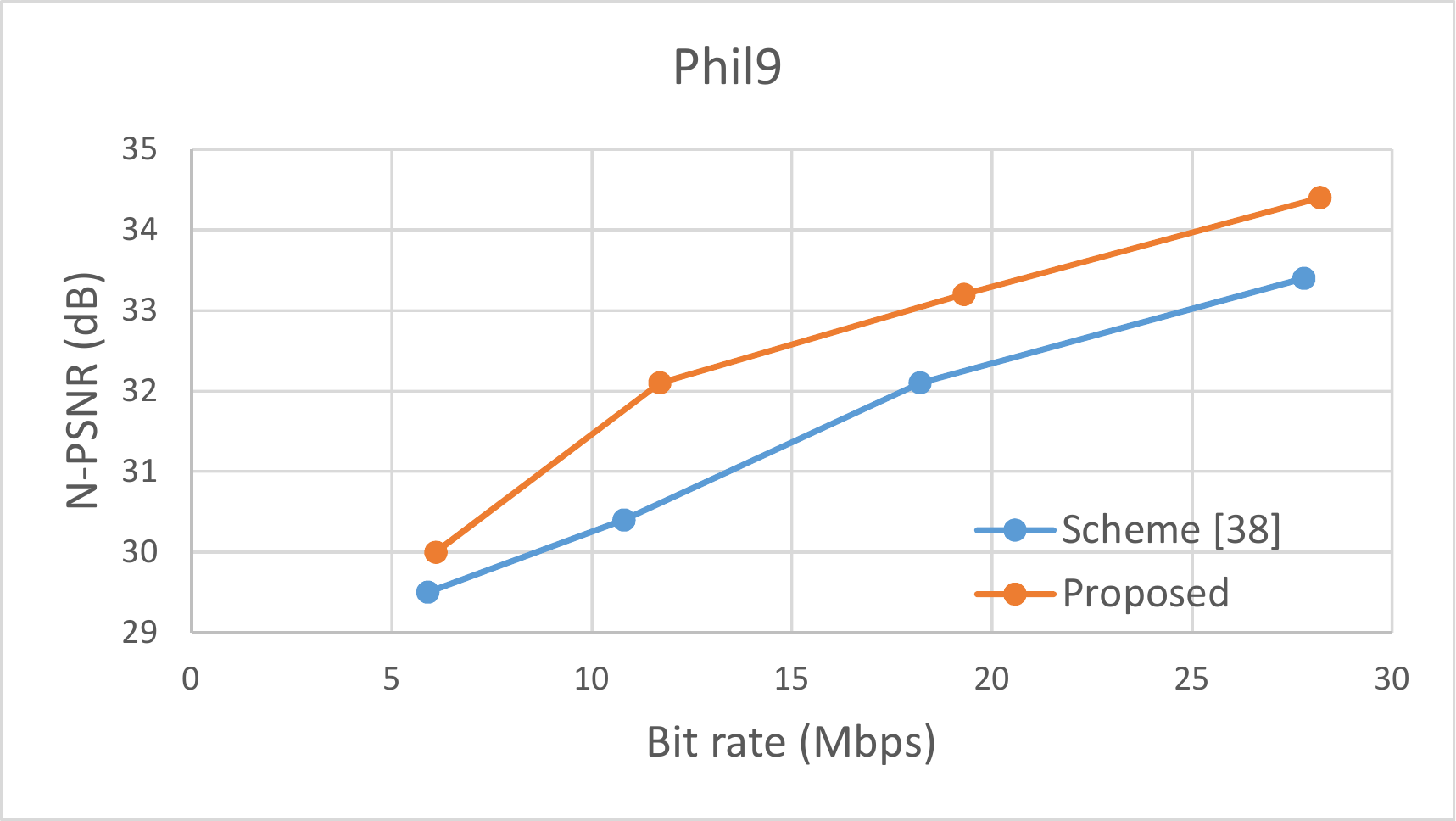}}
		\caption{N-PSNR versus achieved bit rate of the proposed algorithm and scheme \cite{model-based-scheme}. (a) Longdress. (b)  Soldier. (c) Loot. (d) Phil9.}
		\label{visual_RD_NPSNR}
		\vspace{-5mm}
	\end{figure}

{
In addition to using the proposed PC-PSNR to assess the performance, we also compare these two algorithms following the recommendation in \cite{CTC-VPCC}. That is, we compare the achieved bit rate, the D1/D2 geometry PSNR, and { color PSNR (C-PSNR)} for several given target bit rates. Table \ref{Table_Rate_distortion_Comparison} presents the rate and PSNR comparison of our proposed algorithm and \cite{model-based-scheme}, where we calculate the values of those quality metrics using the software PC\_error \cite{Dong_Tian}. In this table, for a fair comparison of the proposed algorithm with \cite{model-based-scheme} on the overall quality of reconstructed point cloud,  we also employ the Normalized PSNR (N-PSNR) metric developed in \cite{model-based-scheme} to measure the quality performance. The N-PSNR is computed by converting the combination of  normalized geometry distortion and color distortion originally in MSE to PSNR measure. As can be observed from the table, our proposed algorithm outperforms \cite{model-based-scheme} consistently in the D1/D2 geometry PSNR, and for some low target bit rate case, the C-PSNR performance of our algorithm may be a little inferior to that of \cite{model-based-scheme}. Nevertheless, our algorithm yields consistent and significant  N-PSNR performance gain for all considered bit rates.  
{Table \ref{QP_selected_by_algorithms} provides examples of the quantization parameters computed by the scheme in \cite{model-based-scheme} and the proposed scheme.}
Fig. \ref{visual_RD_NPSNR} shows the rate-distortion curve comparisons in terms of N-PSNR versus achieved bit rate, and the BD-PSNR \cite{BD-PSNR} between these two curves are provided in the last column in Table  \ref{Table_Rate_distortion_Comparison}. Both quantitative results demonstrate clearly the advantage of our proposed algorithm over \cite{model-based-scheme}.}  

	\begin{table*}[!tbp]
	
	\caption{Rate and PSNR performance of  our proposed algorithm and  \cite{model-based-scheme}. N-PSNR is computed from the weighted average of  D1 geometry distortion and color distortion, both of which are normalized to [0,1]. }
	\centering
	\begin{tabular}{l|l|llll|llll|l}
		\hline
		\multicolumn{1}{c|}{\multirow{2}{*}{Point cloud}} &\multicolumn{1}{c|}{\multirow{2}{*}{\shortstack[l]{Target \\ bit rate \\(Mbps)}}} & \multicolumn{4}{c|}{Scheme  \cite{model-based-scheme}} & \multicolumn{4}{c|}{Proposed}& \multicolumn{1}{c}{\multirow{2}{*}{BD-PSNR}} \\ \cline{3-10}
		& & \shortstack[l]{Bit rate\\(Mbps)}  & D1/D2-PSNR & C-PSNR &N-PSNR & \shortstack[l]{Bit rate\\(Mbps)}  & D1/D2-PSNR & C-PSNR &N-PSNR  \\ \hline
		\multirow{4}{*}{Longdress} &11.6 &     10.9   &    24.3/28.6&    45.6&    35.1&    11.2&    25.8/29.4 & 45.3&    35.6&   \multirow{4}{*}{0.45} \\ \cline{2-10}
		&18.9      & 17.2 &    26.2/31.4&    47.4&    36.9&    18.1&   27.3/32.4&    47.7& 37.5   \\ \cline{2-10}
		&25.5     & 24.2 &    27.4/31.6&    48.1&    37.8&    25.2&    29.2/33.4&    47.4& 38.3   \\ \cline{2-10}
		&32.8      & 31.4 &    27.9/32.1&    48.5&    38.1&    32.3&    29.2/33.4&    48.4& 38.8   \\ \cline{1-11}
		
		\multirow{4}{*}{Soldier} 
		&3.5 &     3.0   &    25.1/30.3&    45.9&    35.4&    3.2&    26.4/31.6 & 45.1& 35.7&   \multirow{4}{*}{0.62} \\ \cline{2-10}
		&7.8      & 6.8 &    26.8/30.9&    46.6&    36.7&    7.3&   28.7/32.9&    46.4& 37.5   \\ \cline{2-10}
		&15.6     & 14.2 &    27.8/31.6&    47.4&    37.6&    15.1&    28.7/32.9&    47.8& 38.4   \\ \cline{2-10}
		&22.4      & 20.9 &    28.6/31.2&    48.2&    38.4&    21.7&    29.4/33.1&       48.4& 39.1   \\ \cline{1-11}
		
		\multirow{4}{*}{Loot} 
		&3.5 &     2.9   &    24.6/28.8&    44.9&    34.7&    3.2&    27.2/31.4 & 44.2&  35.6&   \multirow{4}{*}{0.75} \\ \cline{2-10}
		&7.8      & 7.1 &    25.1/30.4&    45.9&    35.4&    7.5&   27.2/32.4&    45.4& 36.2   \\ \cline{2-10}
		&15.6     & 14.7 &    26.4/31.6&    46.7&    36.5&    15.1&   27.2/32.4&    47.1& 37.2   \\ \cline{2-10}
		&22.4     & 20.7 &    27.1/32.9&    47.1&    37.1&    21.8&    28.2/33.8&    47.3&    37.9   \\ \cline{1-11}
		
		\multirow{4}{*}{Phil9} 
		&6.7 &     5.9  &    22.7/26.5&    40.1&    29.5&    6.1&    24.2/28.9 &    39.7&    30.0&   \multirow{4}{*}{1.15} \\ \cline{2-10}
		&12.1      & 10.8 &    23.3/27.2&    41.4&    30.4&    11.7&   27.4/31.6&    40.8&    32.1   \\ \cline{2-10}
		&19.8     & 18.2 &    26.2/30.0&    42.0&    32.1&    19.3&    27.4/31.6&    42.8&     33.2   \\ \cline{2-10}
		&28.6      & 27.8 &    27.1/31.4&    43.8&    33.4&    28.2&    28.5/32.7&       44.4&     34.4  \\ \cline{1-11}
		
	\end{tabular}\label{Table_Rate_distortion_Comparison}
\end{table*}

\subsection{Time Complexity Analysis and Comparison}
{
The time complexity of the proposed rate-distortion optimization method mainly consists of two parts. One is  the time spent on pre-encoding to determine the parameters of the proposed distortion and rate models and the other is the time induced by the proposed augmented Lagrangian method. In general, the time complexity of the augmented Lagrangian method is much smaller than the pre-encoding procedure, which is thus usually neglected. During pre-encoding, since we need nine pre-encodings to determine the model parameter, while \cite{model-based-scheme} needs three pre-encodings due to its less parameters, our time complexity is almost three times larger than that of \cite{model-based-scheme}. Table \ref{Complexity_Comparison} shows the time complexity comparison between our proposed method with \cite{model-based-scheme}. We run the experiment on a laptop equipped with 3.4G Hz Intel Core i7 Processor and 16 GB RAM, and measure the total running time including the pre-encoding time and bit rate constrained optimization time. } 
	
\begin{table}[!htbp]
	
	\caption{Comparison of the QPs computed by the scheme in \cite{model-based-scheme} and the proposed scheme.}
	\centering
	\begin{tabular}{l|l|cc|cc}
		\hline
		\multicolumn{1}{c|}{\multirow{2}{*}{Point cloud}} &\multicolumn{1}{c|}{\multirow{2}{*}{\shortstack[l]{Target \\ bit rate\\ (Mbps)}}} & \multicolumn{2}{c|}{Scheme  \cite{model-based-scheme}} & \multicolumn{2}{c}{Proposed} \\ \cline{3-6}
		& & $q_g$  & $q_c$ & $q_g$ &$q_c$   \\ \hline
		\multirow{4}{*}{Longdress} &11.6 &     36   &    40&    34&    42     \\ \cline{2-6}
		&18.9      & 35 &    39&    32&    38      \\ \cline{2-6}
		&25.5     & 33 &    37&    28&    37       \\ \cline{2-6}
		&32.8      & 31 &    34&    28&    35     \\ \cline{1-6}	
		
		\multirow{4}{*}{Loot} 
		&3.5 &     34   &    39&    30&    41      \\ \cline{2-6}
		&7.8      & 33 &    38&    30&    40      \\ \cline{2-6}
		&15.6     & 31 &    37&    30&    36       \\ \cline{2-6}
		&22.4     & 30 &    36&    26&    34     \\ \cline{1-6}	
	\end{tabular}\label{QP_selected_by_algorithms}
\end{table}

\vspace{-2mm}
\section{Conclusion}\label{conclusion}
In this paper, we propose a rate-distortion optimization approach for point cloud compression under a given bit rate. Firstly, a unified distortion model is proposed, which computes the overall distortion of the point clouds from the geometry distortion and color distortion by considering the correlation between geometry and color variables. Then, we approximate the relationships of the overall distortion and bit rate with the geometry and color quantization parameters using fitted polynomials. Finally, we formulate the bit rate constrained coding problem for point clouds, and solve it using the augmented Lagrangian method. Experimental results demonstrate that the overall distortion of point clouds is much more sensitive to the bit rate change of geometry coding than that of color coding, and the proposed rate distortion optimization algorithm can exploit this fact to choose the optimal color and geometry quantization parameters. Experiments also verify that the proposed algorithm  outperforms other schemes significantly in terms of objective and subjective qualities.  

\begin{table}[htbp]
	\centering
	\caption{Time complexity comparison between  our proposed algorithm and  \cite{model-based-scheme}.}
	\begin{tabular}{l|l|l}
		\hline
		\multicolumn{1}{c|}{\multirow{2}{*}{Point cloud}} & \multicolumn{2}{c}{Encoding time (s)} \\ \cline{2-3}
		& Scheme  \cite{model-based-scheme}  & Proposed \\ \hline
		Longdress & 4527.65                                           &    13567.21                                         \\ \hline
		Loot      & 5486.31                                           &    16349.38                                      \\ \hline
		Soldier   & 8542.68                                           & 25694.50                                         \\ \hline
		Richardo10     & 2346.41                                          & 7042.57                                          \\ \hline
		Phil9   & 2578.91                                          & 7711.62                                         \\ \hline
	\end{tabular}\label{Complexity_Comparison}
\end{table}

\section*{Acknowledgement}
The authors would like to thank Prof. A. Smolic from Trinity College Dublin for the guidance and comments on various technical issues examined in this paper, and Dr. E. Zerman from Mid Sweden University for the help on the validation of  the  unified distortion model proposed in this paper.

\end{document}